\documentclass[12pt]{article}

\usepackage{scicite}

\usepackage{times}

\usepackage{amssymb}
\usepackage{graphicx}
\usepackage{caption}
\usepackage{xurl}
\newcommand{\kms}{\ensuremath{\, \mathrm{km}\, {\mathrm s}^{-1}}}
\newcommand{\cms}{\ensuremath{\, \mathrm{cm}\, {\mathrm s}^{-2}}}
\newcommand{\smy}{\,$M_\odot$\,${\rm yr}^{-1}$}

\newenvironment{sciabstract}{%
\begin{quote} \bf}
{\end{quote}}

\title{\vspace{-2cm} A massive helium star with a sufficiently strong magnetic field to form a magnetar} 

\author{
Tomer Shenar$^{1}$,
Gregg A. Wade$^{2}$,
Pablo Marchant$^{3}$,
Stefano Bagnulo${^4}$\\
Julia Bodensteiner$^{5, 3}$, 
Dominic M. Bowman$^{3}$,  
Avishai Gilkis$^{6}$, 
Norbert Langer$^{7, 8}$, \\
Andr\'e Nicolas-Chen\'e$^{9}$, 
Lidia Oskinova$^{10}$,
Timothy Van Reeth$^{3}$, 
Hugues Sana$^{3}$,\\
Nicole St-Louis$^{11}$,
Alexandre Soares de Oliveira$^{12}$,
Helge  Todt$^{10}$,
Silvia Toonen$^{1}$\\
\scriptsize{$^{1}$Anton Pannekoek Institute for Astronomy, University of Amsterdam, Amsterdam 1098 XH , the Netherlands}\\
\scriptsize{$^{2}$ Department of Physics and Space Science, Royal Military College of Canada, Kingston K7K7B4, Canada}
\\
\scriptsize{$^{3}$ Institute of Astronomy, Katholieke Universiteit (KU) Leuven, Leuven 3001, Belgium}
\\
\scriptsize{$^{4}$ Armagh Observatory \& Planetarium, College Hill, Armagh BT61 9DG, UK}
\\
\scriptsize{$^{5}$ European Southern Observatory,  Garching bei M\"unchen 85748, Germany}
\\
\scriptsize{$^{6}$ The School of Physics and Astronomy, Tel Aviv University, Tel Aviv 6997801, Israel}
\\
\scriptsize{$^{7}$ Argelander-Institut f\"ur Astronomie, Universit\"at Bonn, Bonn 53121, Germany} 
\\
\scriptsize{$^{8}$ Max-Planck-Institut für Radioastronomie,  Bonn 53121, Germany} 
\\ 
\scriptsize{$^{9}$ National Science Foundation's National Optical-Infrared Astronomy Research Laboratory (NSF's NOIRLab), Hawai`i 96720, USA}
\\
\scriptsize{$^{10}$ Institut f\"{u}r Physik und Astronomie, Universit\"{a}t Potsdam, Potsdam D-14476, Germany}
\\
\scriptsize{$^{11}$ D\'epartement de physique, Universit\'e de Montr\'eal, Complexe des sciences, Montr\'eal H2V 0B3, Canada}\\
\scriptsize{$^{12}$ Institute of Research and Development, Universidade do Vale do Para\'\i ba, 12244-000, S\~ao Jos\'e dos Campos, Brazil}
\\
\scriptsize{$^\ast$To whom correspondence should be addressed; E-mail:  T.Shenar@uva.nl}}
\date{\today}

\date{}

\begin{document} 


\baselineskip24pt


\maketitle


\begin{sciabstract}
Magnetars are highly magnetized neutron stars; their formation mechanism is unknown. Hot helium-rich stars with spectra dominated by emission lines are known as Wolf-Rayet stars. We observe the binary system HD~45166 using spectropolarimetry, finding that it contains a Wolf-Rayet star with a mass of 2 solar masses and a magnetic field of 43 kilogauss. Stellar evolution calculations indicate that this component will explode as a type Ib or IIb supernova, and the strong magnetic field favors a
magnetar remnant. We propose that the magnatized Wolf-Rayet star formed by the merger of two lower mass helium stars. 
\end{sciabstract}

% In setting up this template for *Science* papers, we've used both
% the \section* command and the \paragraph* command for topical
% divisions.  Which you use will of course depend on the type of paper
% you're writing.  Review Articles tend to have displayed headings, for
% which \section* is more appropriate; Research Articles, when they have
% formal topical divisions at all, tend to signal them with bold text
% that runs into the paragraph, for which \paragraph* is the right
% choice.  Either way, use the asterisk (*) modifier, as shown, to

% \section*{Introduction}

Neutron stars form in supernovae, by the collapse of stellar cores that exceed the Chandrasekhar mass limit [mass $M \gtrsim 1.4$ solar masses ($\,M_\odot$)]. 
Bare stellar cores can be exposed as hot, evolved helium-rich stars that have shed their outer hydrogen-rich layers. A subset of these massive helium stars are observed as Wolf-Rayet stars, which have spectra dominated by broad emission lines produced by strong  stellar winds \cite{Crowther2007, Langer2012}.  Massive helium stars (i.e. $M \gtrsim 1.4\,M_\odot$)  are thought to be stripped products of massive stars that lost their hydrogen-rich envelopes through stellar winds, eruptions, or interactions with a binary companion \cite{Paczynski1967, Woosley2019}. Alternatively, massive helium stars may be produced through the merger of lower-mass objects \cite{Zapartas2017}. 
% Virtually all known massive helium stars exhibit current masses larger than $\approx 8\,M_\odot$ and are classified as Wolf-Rayet stars: hot stars with emission-line dominated spectra. HD~45166, the subject of our study, is a notable exception (see below).

Roughly $10\%$ of young neutron stars have magnetic fields $> 10^{14}$\,gauss (G) \cite{Kaspi2017}. These are known as magnetars; their origin is  debated \cite{Mereghetti2015, Ferrario2015}. One formation scenario invokes fossil magnetic fields  rooted in the pre-collapse massive core \cite{Ferrario2006}. About 7-10\% of massive main-sequence stars have strong (several kG) large-scale surface magnetic fields \cite{Wade2016,Grunhut2017}; these could be progenitors of magnetars. However, corresponding magnetic fields have not been detected in evolved massive stars \cite{delaChevrotiere2014}. Strongly magnetic low-mass helium stars have been observed \cite{Dorsch2022,Pelisoli2022, Shultz2021}, but not massive magnetic helium stars (i.e.\ exceeding the Chandrasekhar mass limit).
% , with the exception of ambiguous reports of weak magnetic fields \cite{Hubrig2020}.  

% Massive stellar cores  are typically thought to form in the interiors of stars born with at least eight solar masses ($M_{\rm ini}$ \gtrsim 8\,$M_\odot$). Alternatively, they may be produced through the coalescence of lower-mass stellar objects. 
% After exhausting hydrogen in their cores, massive stars tend to expend and become cool ($T_{\rm eff} \lesssim 4\,$kK), bloated red supergiants before ending their lives in a powerful supernova event. However,

% Almost all massive helium stars currently known have masses in excess of $7-8\,M_\odot$, and exhibit emission-line dominated spectra that are the defining feature of so-called Wolf-Rayet (WR) stars \cite{Shenar2020LB1, nuGem...}.

% Here, we report on the detection of a strong magnetic field in  the helium-star component of HD~45166, totalling to a modulus of $|B| = 43\pm 3\,$kG . 

\begin{figure}[!htb]
   \centering
\includegraphics[width=\textwidth]{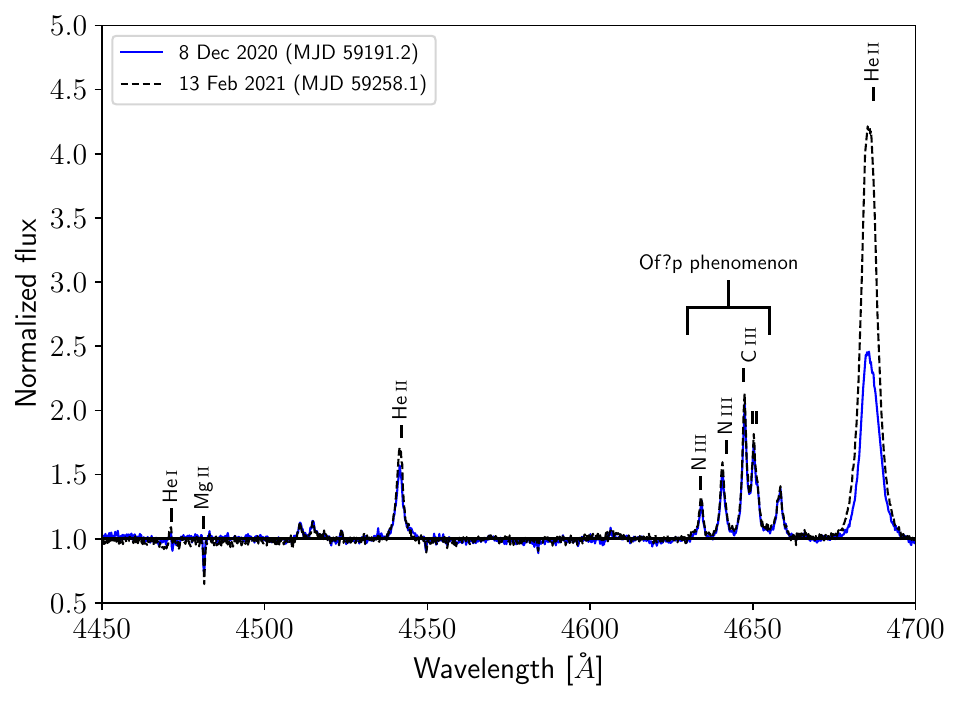}
    \caption{{\bf Indications for magnetism in HD~45166}. Two optical HERMES spectra of HD~45166 were taken $\approx 70$\,d apart (see legend). The presence of the so-called Of?p phenomenon \cite{Walborn1972, Naze2010} and the strong variability observed in the He\,{\scshape ii}\,$\lambda 4686$ line are common features of hot magnetic stars, indicating that the Wolf-Rayet component might be magnetic. The narrow Mg\,{\scshape ii}\,$\lambda 4481$ line is due to the B7~V component.
    } 
    \label{fig:Spec}
\end{figure}

The HD~45166 (ALS~8946) system is a binary comprising a main sequence star (classified as spectral type B7~V) with a hot stellar companion. The hot companion has a spectrum dominated by the characteristic emission lines of a Wolf-Rayet star (Fig.\ \ref{fig:Spec}), but it was classified as ``quasi Wolf-Rayet'' (qWR) star owing to its peculiarly narrow emission lines (widths of hundreds of \kms~instead of thousands), spectral variability, and the anomalous presence of strong carbon, oxygen, and nitrogen lines in its spectrum.
Previous radial velocity (RV) measurements have shown a 1.6\,d periodicity in the velocity of the B7~V component, interpreted as the orbital period of the system \cite{Steiner2005}. This implied a mass of $4.2\pm 0.7\,M_\odot$ for the Wolf-Rayet component and a pole-on orbital configuration (inclination $i=0.7^\circ$) \cite{Steiner2005}. That mass is well below the typical masses of Wolf-Rayet stars in our Galaxy [$M \gtrsim 8\,M_\odot$, \cite{Hamann2019}]; no other massive helium stars are known with $M\lesssim 8\,M_\odot $ \cite{Goetberg2018}. Comparison to stellar atmosphere models that do not assume local thermodynamic equilibrium (non-LTE) found that the Wolf-Rayet component is a hot (surface effective temperature of  $T_* = 70\,$kK), helium-rich star with enhanced nitrogen and carbon contents compared to a solar composition \cite{Groh2008}. A latitude-dependent wind model was invoked to reproduce the spectrum of the Wolf-Rayet component \cite{Groh2008}.  The implied mass-loss rate is orders of magnitude higher than predictions for helium stars of this mass \cite{Vink2017,Sander2020,Shenar2020_WR}.

% \begin{figure}
%   \centering
% \includegraphics[width=0.8\textwidth]{Figs/Sketch_qWR.jpeg}
%     \caption{ Visualisation of HD~45166. The system is found to comprise a hot, helium-rich, strongly magnetised $2.1\,M_\odot$ star bound to a  pulsating $3.5\,M_\odot$ B7~V star. While an orbital period of 1.6\,d was proposed in the literature, we find that the true orbital period is of the order of decades. The sizes in the image are not to scale (Image credit: Fabian Bodensteiner).
%     } 
%     \label{fig:visu}
% \end{figure}

The presence of a C and N emission-line complex in the range $4630-4660\,$\AA~(referred to as the Of?p phenomenon) and strong spectral variability are typical signatures of magnetism in hot stars \cite{Walborn1972, Naze2010}. Here, we use spectropolarimetry to investigate whether the Wolf-Rayet component in HD~45166 is magnetic.
% Here we use spectropolarimetry to show that the Wolf-Rayet component
% {\bf comprises a long-period binary with the B7~V component, and}  
% exhibits a magnetic field that is stronger than any magnetic field ever measured for a non-degenerate super-Chandrasekhar mass object. 

\section*{Observations of HD~45166}

We collected eight high-resolution spectropolarimetric observations (spectra of Stokes parameters $I$ and $V$)  of HD~45166 in February 2022 with the ESPaDOnS spectropolarimeter at the Canada-France-Hawaii Telescope (CFHT) \cite{MaterialsMethods}. The spectra cover the range \mbox{3668 - 10\,480\,\AA} and are  used to measure the magnetic-field strength in the Wolf-Rayet and B7~V components of HD~45166.  We additionally use spectra acquired with three spectrographs for RV monitoring \cite{MaterialsMethods}. 103 spectra were obtained with the Coud\'e spectrograph at the
1.6\,m telescope of Laborat\'orio Nacional de Astrofísica (LNA), mainly covering the range 4520 -- 4960\,\AA. 36 spectra were obtained with the  Fiber-fed Extended Range Optical Spectrograph (FEROS) at the 1.52\,m telescope of the European Southern Observatory (ESO), covering the range 3830 -- 9215\,\AA. Finally, 28 spectra were acquired with the {\scshape HERMES} spectrograph mounted on the 1.2\,m Mercator telescope, covering the range 3770--9000\,\AA. The analysis of the spectrum and spectral energy distribution (SED) relies on additional ultraviolet (UV) and photometric data \cite{MaterialsMethods}.

% on  t analysis and  For the spectral analysis, we append these data with ultraviolet (UV) spectra acquired  with the International Ultraviolet Explorer (IUE) in the ranges 1150-2150\,\AA\ and 1850-3350\,\AA, and with the Far Ultraviolet Spectroscopic Explorer (FUSE) in the range 900--1200\AA. For the analysis of the spectral energy distribution (SED), We also use photometry in the $UBVJHK$ bands and infrared (IR) photometry from the Wide-field Infrared Survey Explorer (WISE) \cite{MaterialsMethods}.

% (FEROS) at the 1.52\,m telescope of the European Southern Observatory (ESO) in La Silla, Chile, covering the range $3830 - 9215\,$\AA\ at a resolving power of $R = 48\,000$ \cite{MaterialsMethods}.Finally, 28 spectra were acquired with the {\scshape HERMES} spectrograph mounted on the 1.2\,m Mercator telescope at the Observatorio del Roque de Los Muchachos on La Palma, Spain \cite{MaterialsMethods}. For the spectral analysis, we append these data with ultraviolet (UV) spectra acquired  with the International Ultraviolet Explorer (IUE) in the ranges 1150-2150\,\AA\ and $1850-3350\,$\AA\, and with the Far Ultraviolet Spectroscopic Explorer (FUSE) in the range 900-1190\AA\. We append these data with photometry in the $UBVJHK$ bands and infrared (IR)  -$band photometry from \cite{Abazajian2009}, $BV-$band photometry from \cite{Zacharias2004}, $JHK$ photometry from a compilation by \cite{Bonanos2009}, and infrared photometry from the Wide-field Infrared Survey Explorer (WISE) \cite{MaterialsMethods}.

\section*{Evidence for magnetism}

\begin{figure}[!htb]
   \centering
\includegraphics[width=\textwidth]{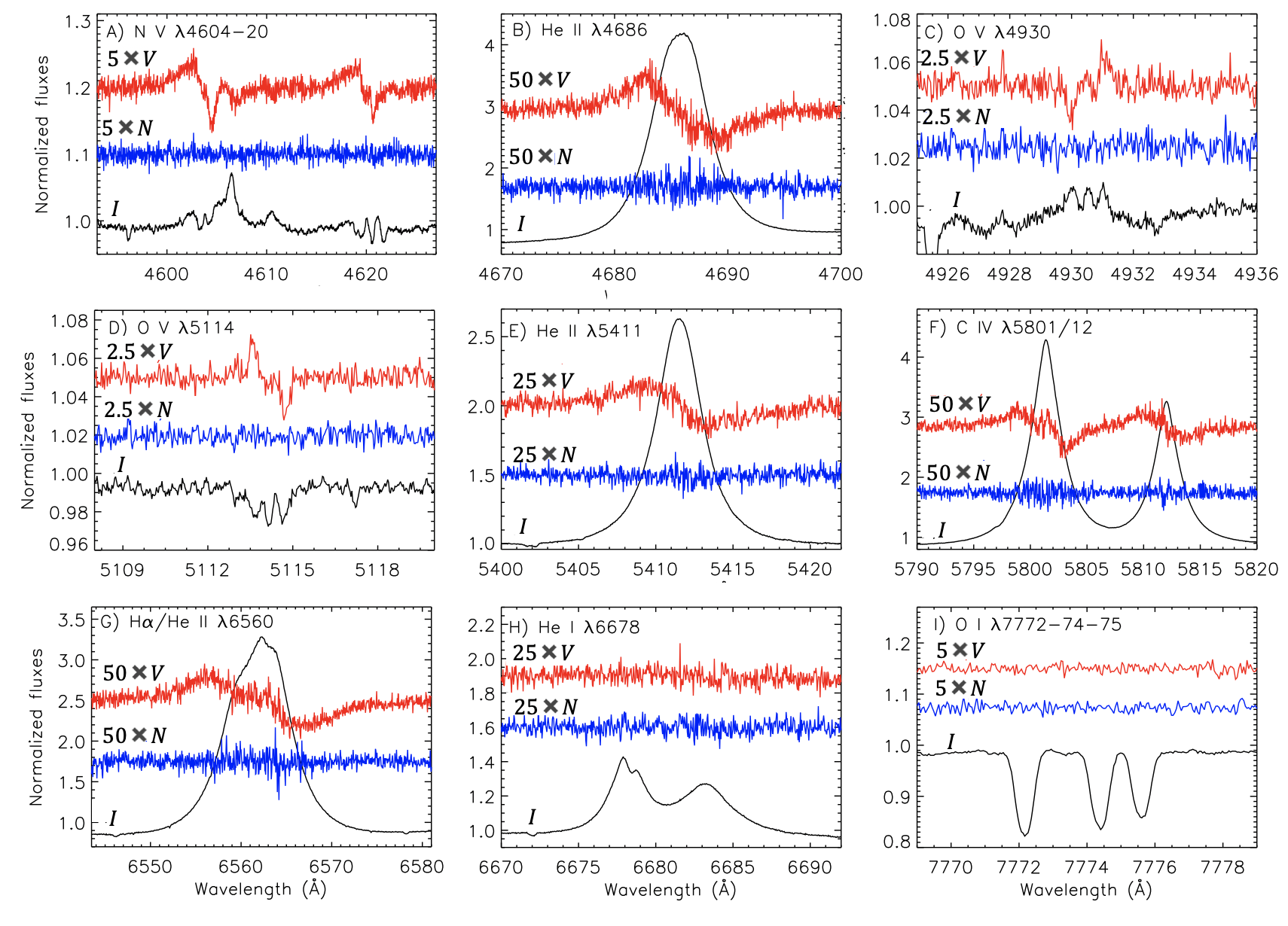}
\vspace{0.5cm}
    \caption{{\bf Stokes spectra of HD~45166.} The panels show the intensity spectrum ($I$, black lines), diagnostic null spectrum ($N$; blue lines), and Stokes $V$ spectrum (upper curves, red lines) of the co-added ESPaDOnS spectrum. The $N$ and $V$ spectra have been vertically shifted and scaled for display purposes (the multiplication factors are listed in each panel). (A) -- (H) Several diagnostic lines of the Wolf-Rayet component. (I)  An O\,{\scshape i} triplet associated with the B7~V star. Zeeman splitting is visible in the 
     O\,{\scshape v}\,$\lambda 4930$ and O\,{\scshape v}\,$\lambda 5114$ lines (C, D). We use these lines to measure the magnetic field strength. There is no Stokes $V$ signature visible for lines associated with the B7~V star (panel I). 
    } 
    \label{fig:Zeeman}
\end{figure}

We detect strong circular polarization (Stokes $V$) in the ESPaDOnS spectra 
% -- unambiguous evidence for the presence of a very strong magnetic field --
in the majority of lines associated with the Wolf-Rayet component (Fig.\,\ref{fig:Zeeman}).   We also detect the magnetic splitting (i.e. Zeeman splitting) of two spectral lines belonging to  O\,{\scshape v} in our spectrum (Fig.\,\ref{fig:Zeeman}C-D) that form in or close to the stellar surface. We measure the magnetic field from the separation of the split Zeeman components of these lines, finding  a mean field modulus $\langle B \rangle_{\rm qWR} = 43.0\pm2.5\,$kG and a mean longitudinal
field  $\langle B_{\rm z} \rangle_{\rm qWR} = 13.5\pm2.5$\,kG (Table\,\ref{tab:Parameters}).
The ratio of the longitudinal field to the modulus ($\approx$1/3rd) is  consistent with a dipolar magnetic field viewed near the magnetic pole \cite{Petit2013}.

\begin{table}[!htb]
\centering
\caption{{\bf Derived parameters for HD~45166.} Provided are the distance, reddening $E_{B-V}$, age, orbital elements (period $P$, time of periastron $t_0$, eccentricity $e$, argument of periastron $\omega$, systemic velocity $v_0$, semi-major axis $a$,  mass ratio $q$, orbital inclination $i$, and RV amplitudes $K_1, K_2$),  mean magnetic-field modulus and line-of-sight components $\langle B \rangle$ and $\langle B_{\rm z} \rangle $,  minimum masses $M \sin^3 i$, masses $M$, relative contributions to the light in the $V$-band $l(V)$, effective surface temperatures $T_*$, luminosities $\log L$, radii $R_*$, projected rotational velocities $v \sin i$, rotational period $P_{\rm rot}$, and elemental abundances of He, C, N, and O ($X_{\rm He}$, $X_{\rm C}$, $X_{\rm N}$ ,$X_{\rm O}$)  in mass fractions.   Uncertainties are 68\% confidence interval ($1\sigma$). }
%\resizebox{.5\textwidth}{!}{
\begin{tabular}{lccc}
\hline  \hline
Parameter & unit & B7~V & Wolf-Rayet  \\
\hline
\vspace{-4mm}\\ 
Distance (adopted) & parsec & \multicolumn{2}{c}{$991^{+38}_{-33}$ \cite{Bailer-Jones2021}}  \\
$E_{B - V}\,$ & magnitude & \multicolumn{2}{c}{$0.210\pm0.010$ } \\
age  & Myr & \multicolumn{2}{c}{$105\pm35$ } \\
$P\,$ & day & \multicolumn{2}{c}{$8200 \pm 190$}   \\
$t_0$ (MJD) &  - &  \multicolumn{2}{c}{ $49820\pm 360$ }   \\
$e$ & - & \multicolumn{2}{c}{$ 0.46\pm 0.18$ }   \\
$\omega$ & degree & \multicolumn{2}{c}{$132 \pm 11$ }     \\
$v_0$ & \kms & \multicolumn{2}{c}{$5.7 \pm 1.8$ }   \\
$a$ & au &  \multicolumn{2}{c}{$10.5 \pm 1.8$ }  \\
$q\equiv M_{\rm qWR} / M_{\rm B}$ & - & \multicolumn{2}{c}{$0.60 \pm 0.13$ } \\
$i$ & degree  & \multicolumn{2}{c}{$49 \pm 11$ }   \\
$K$ & \kms & $5.8 \pm 1.3$ & $9.9 \pm 1.6$ \\
$\langle B \rangle$ & kG & - & $ 43.0\pm2.5$ \\
$\langle  B_{\rm z} \rangle $ & kG  & $<0.02$ & $13.5\pm2.5$   \\
$M \sin^3 i$ & $M_\odot$  & $1.50\pm0.74$ & $0.85\pm0.45$   \\ 
$M$ & $M_\odot$ &$3.40\pm0.06$ &   $2.03\pm0.44$ \\ 
$l (V)$ & - & $0.460\pm0.050$   &  $0.560\pm0.050$ \\
$T_*$ & kK   &   $13.00\pm0.50$  & $56.0\pm6.0$ \\
$\log L/L_\odot$ & - & $2.250 \pm 0.050$ &  $3.830\pm0.050$  \\
$R_*$ & $R_\odot$  & $2.63\pm 0.41$ &  $0.88\pm 0.16$  \\
$v \sin i $ & \kms & $\lesssim 10$  &  $\lesssim 10$   \\
$P_{\rm rot}$ &  day & - &  $124.82\pm0.21$    \\
$X_{\rm He}$  (adopted) & - & 0.25 \cite{Asplund2009} &  $0.67$  \cite{Groh2008} \\
$X_{\rm C}$ (adopted) & - & 0.0024 \cite{Asplund2009} & $0.0059$ \cite{Groh2008}\\
$X_{\rm N}$ (adopted) & - & 0.00069 \cite{Asplund2009}  & $0.0020$ \cite{Groh2008} \\
$X_{\rm O}$ (adopted) & - & 0.0057 \cite{Asplund2009}  &  $0.0015$ \cite{Groh2008} \\
\hline
\end{tabular}
% \\[0.1cm]
\label{tab:Parameters}
\end{table}

Such a strong magnetic field in the Wolf-Rayet component implies that its emission-line spectrum is formed in plasma confined to the magnetic field loops (the magnetosphere), and not in a radially expanding stellar wind  \cite{MaterialsMethods}. In that case, one-dimensional (1D) stellar atmosphere models [as used in previous studies \cite{Groh2008}] are insufficient for analyzing the emission lines, which affected the derived physical parameters. We reanalyze the spectra and SED using the non-LTE Potsdam Wolf-Rayet ({\scshape powr}) model atmosphere code \cite{Hamann2003, Sander2015}. While the {\scshape powr} code is 1D, we only use it to analyze features originating in the stellar surface of the Wolf-Rayet component, ignoring its emission features \cite{MaterialsMethods}. The resulting physical parameters are listed in Table\,\ref{tab:Parameters}. We find a high effective temperature for the Wolf-Rayet component, though $\approx$15kK lower than previously determined \cite{Groh2008}. The effective temperature  ($T_* = 56.0\pm 5.0\,$kK) and bolometric luminosity (in units of solar luminosity $L_\odot$, $\log L/L_\odot = 3.830\pm0.050$) indicate  that the magnetic Wolf-Rayet component is not a main sequence star, but an  evolved object.  We applied the same analysis to the B7~V component. By using its derived stellar parameters as an input in the {\scshape BONNSAI} Bayesian tool \cite{Schneider2014}, we derive $M_{\rm B} = 3.38\pm0.10\,M_\odot$ and $105\pm 35\,$Myr for mass and age of the B7~V component, respectively  \cite{MaterialsMethods}.

\begin{figure}[!htb]
   \centering
\begin{tabular}{cc}
\includegraphics[width=0.51\textwidth]{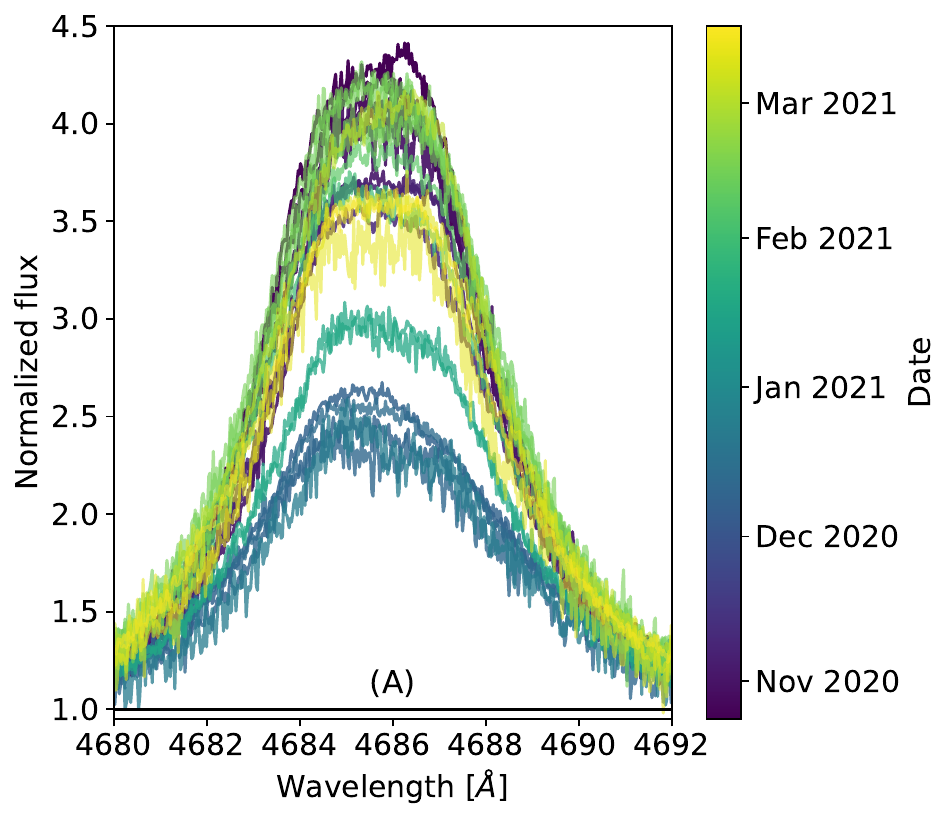}&
\includegraphics[width=0.44\textwidth]{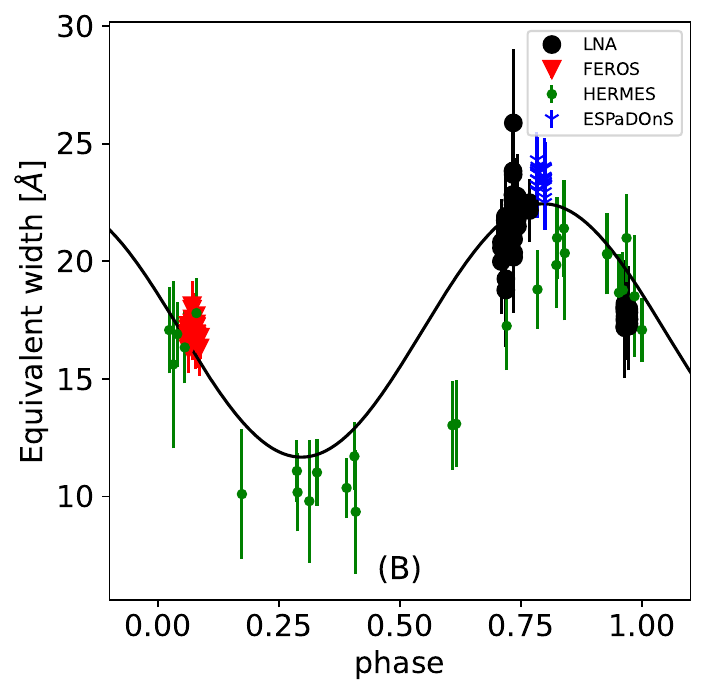}
\end{tabular}
    \caption{{\bf Variability of the He\,{\scshape ii}\,$\lambda 4686$ emission line.}  (A) HERMES spectra of the  He\,{\scshape ii}\,$\lambda 4686$ line (color depicts the time of observation), illustrating the changing strength of this line on a timescale of weeks. (B) Line strengths (equivalent widths) of He\,{\scshape ii}\,$\lambda 4686$  over the 24-yr spectroscopic dataset. Different colors and plotting symbols indicate different spectrographs, as indicated in the legend. The measurements were phased at the derived period of 124.8\,d. We interpret this as the rotational period of the Wolf-Rayet component. Error bars are estimated 1$\sigma$ confidence intervals.} \label{fig:EW}
     % I can do without the label too
\end{figure}

The Wolf-Rayet component exhibits changes in the strength of lines such as He\,{\scshape ii} $\lambda 4686$. These appear to be periodic (Fig.\,\ref{fig:EW}A), with a best-fitting period of $124.8\,$d (Fig.\,\ref{fig:EW}B). Periodic changes in the line strengths of hot magnetic stars are typically interpreted as their rotational periods \cite{Donati2006, Wade2011}, implying a rotational period $P_{\rm rot, qWR} = 124.8\pm0.2\,$d for the Wolf-Rayet component.  This period is consistent with the projected rotational velocity  indicated by the narrow O\,{\sc v} lines (projected rotational velocity $v \sin i \lesssim 10\,\kms$).

\section*{The orbit of HD~45166}

The ESPaDOnS spectra indicate that the B7~V component exhibits spectral line profile variations caused by non-radial gravity-mode pulsations (Fig.\,\ref{fig:Pulsations}). This is supported by a frequency analysis of a light curve obtained with the Transiting Exoplanet Survey Satellite (TESS) performed in this study \cite{MaterialsMethods}, which yields 1.6\,d as one of the significant periods.
we conclude that the 1.6-d period previously attributed to orbital motion \cite{Steiner2005} is caused by pulsations in the B7~V component.  We therefore need to reassess the orbit of  and the mass of the Wolf-Rayet component, $M_{\rm qWR}$.

\begin{figure}[!htb]
   \centering
\includegraphics[width=\textwidth]{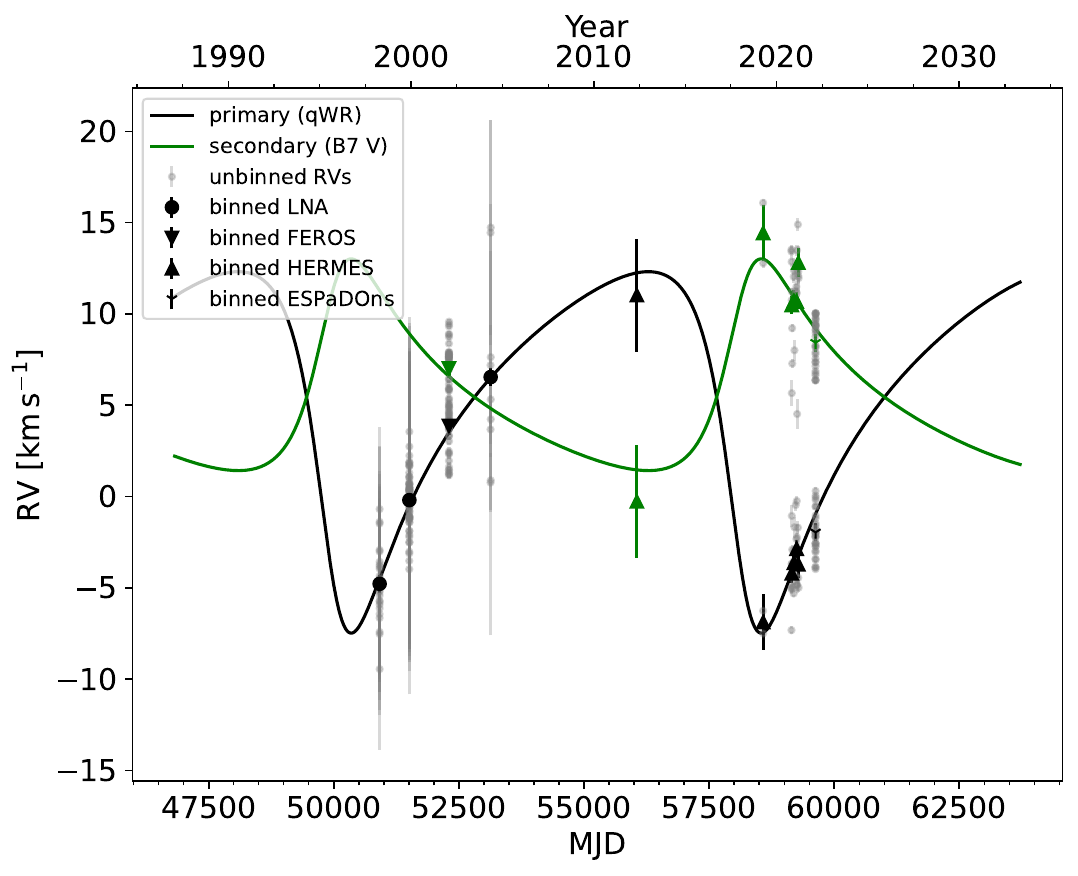}
    \caption{{\bf Two-component orbital solution of HD~45166}. The individual RVs of the Wolf-Rayet and B7~V components (gray symbols) are binned by 30\,d (green and black symbols for Wolf-Rayet and B7~V components, respectively). The different instruments are shown as distinct symbols (see legend).  Solid lines correspond to the best-fitting RV curves for the Wolf-Rayet (black) and B7~V (green) components. The fit yields a reduced $\chi^2$ of 1.21. Error bars indicate  estimated 1$\sigma$ confidence intervals \cite{MaterialsMethods}.
    } 
    \label{fig:Orbit}
\end{figure}

To determine the orbit, we compiled spectroscopic data spanning 24 years taken with the facilities listed above \cite{MaterialsMethods}. These data show a long-term anti-phase motion of the B7~V and Wolf-Rayet components (Fig.\,\ref{fig:Orbit}). We find multiple periodicities associated with both the Wolf-Rayet and B7~V  components \cite{MaterialsMethods}, but we determine a best-fitting orbit (Table\,\ref{tab:Parameters}). 
We find an orbital period $P=8200\pm190\,$d and a semi-major axis $a = 10.5\pm1.8$ astronomical units (au).  This indicates that the components are much more widely separated than had previously been determined. Our derived  mass ratio of  $q \equiv M_{\rm qWR}/M_{\rm B} = 0.60\pm0.13$, combined with the  mass of the B7~V component  ($M_{\rm B} = 3.38\pm0.10\,M_\odot$) we derived above, implies $M_{\rm qWR} = 2.03\pm0.44\,M_\odot$ (uncertainties are 68\% confidence intervals). 
This is less than the $4.2\,M_\odot$ previously reported, but still above the Chandrasekhar limit. 
The derived mass and luminosity are consistent with mass-luminosity relations for He stars \cite{Graefener2011}. The large separation and implied orbital inclination of $i = 49\pm 11^\circ$ (derived from $M_{\rm B} \sin^3 i$ and $M_{\rm B}$, Table\,\ref{tab:Parameters}) explain the low-amplitude RV motion of both binary components, without invoking a pole-on configuration.

\section*{Implications on the magnetar population}

With a mass of $2.03\pm0.44\,M_\odot$, we can expect the Wolf-Rayet component to evolve until it collapses into a neutron star. This is supported by evolution models constructed for the system, although the final fate of the Wolf-Rayet component depends on modeling uncertainties \cite{MaterialsMethods}. Upon core-collapse, magnetic flux conservation leads to an increase in the magnetic field at the surface.  With a stellar radius of $R_{*, \rm qWR} = 0.88\pm0.16$ solar radii ($R_\odot$) calculated via the Stefan-Boltzmann relation, $\langle B \rangle_{\rm qWR} = 43.0\pm 2.5\,$kG, and assuming a final neutron-star radius of $12\,$km \cite{Pang2021}, we calculate a final magnetic field of the neutron star $\langle B \rangle_{\rm NS} = (1.11 \pm 0.42) \times 10^{14}$\,G. This is within the range observed for magnetars  [$\langle B \rangle \gtrsim 10^{14}$\,G, \cite{Duncan1992}]. Our measurements and evolution models therefore indicate that the Wolf-Rayet component in  is an immediate progenitor of a magnetar.

All magnetars in the Milky Way are isolated i.e.\ do not have a binary companion \cite{Kaspi2017}. For the Wolf-Rayet component in HD~45166, we expect the mass-loss and velocity kick imparted on the magnetar by the supernova explosion to disrupt the system, given the large orbital separation. With an estimated rotation period of 125\,d and an estimated radius of $\approx 0.3\,R_\odot$ for the helium core of the Wolf-Rayet component \cite{Langer1989}, angular momentum conservation implies that the magnetar immediately after collapse would have a spin period $\lesssim 40\,$ms. This is similar to the spin period of the famous Crab Pulsar ($33\,$ms), which is a neutron star formed about 1000 years ago. Such a spin is not expected to provide sufficient energy to power a superluminous supernova or a long duration gamma-ray burst \cite{Woosley2010, Blanchard2020}. 

% Instead, HD~45166 provides evidence that Milky Way magnetars are formed born slowly rotating \cite{Martin2014}, and that their strong magnetic field existed already before the core collapse of their progenitor stars.

% Magnetars are always observed in isolation \cite{Kaspi2017}. This will likely be the fate of the Wolf-Rayet component in , too. At collapse, the mass-loss and kick imparted on the newly-born magnetar due to the supernova explosion is expected to disrupt the system, especially considering the very large measured separation between the components. The collapse accompanying the formation of the magnetar is expected to result in a magnetar-powered supernova, which is a favoured model for explaining a subset of luminous supernovae observed across the Cosmos up to high redshifts \cite{Kasen2010}.

\section*{Evolutionary history of the system}

We next consider how the Wolf-Rayet component itself formed. We exclude the possibly that it is the stripped descendant of a massive star, because to produce a $2\,M_\odot$ stripped core, the progenitor star would need to have had an initial mass of  $\approx 10\,M_\odot$. Single-star evolution models do not predict that stars of that mass strip themselves \cite{Maeder2000, Shenar2020_WR}, and the B7~V companion is too far away for binary interactions to have stripped it. In addition,  the total lifetime (including post main-sequence evolution) of a $10\,M_\odot$ star would be $\approx$ 30\,Myr \cite{Brott2011}, well below the derived age of the B7~V component ($105\pm35\,$Myr), so we reject the possibility that they could both be present in the same binary system.

Stellar mergers have been proposed as a potential origin of  magnetic stars \cite{Ferrario2009, Wickramasinghe2014, Schneider2019, Bagnulo2022}. Strong magnetic fields have been identified in low-mass helium stars (OB-type subdwarfs) and were suggested to originate in merger events between two white dwarfs \cite{Dorsch2022,Pelisoli2022}. However,  to produce a $\approx 2\,M_\odot$ helium star through a merger of white dwarfs, the merger would need to have involved rare, massive CO or ONe white dwarfs. Models predict that such a merger product would to either immediately explode as a supernova \cite{Webbink1984}, or do so after a short lifespan  of approximately 10\,kyr \cite{Schwab2016}, suggesting that observing a stellar product following the merger is unlikely.
% Moreover, the presence of  hydrogen and nitrogen in the spectrum of the Wolf-Rayet component \cite{Groh2008} is difficult to reconcile with such a scenario.
% the white dwarfs would need to be rather massive ($M \gtrsim 1\,M_\odot$; see Supplementary Materials). While such a merger is expected to result in may avoid a thermonuclear explosion \cite{Schwab2016, Gvaramadze2019}. On the other hand, the  B7~V companion effectively provides us with a clock. With an estimated age of $105\pm35\,$Myr, forming two white dwarfs and merging them within this timescale is not possible in the framework of standard binary evolution \textcolor{red}{[Pablo/Avishai/Norbert: exaggeration?}], disregarding the unlikely event that the qWR+B7~V binary became bound only after the merger event. The presence of hydrogen in the atmosphere of the Wolf-Rayet component also disfavours a white-dwarf merger model.

\begin{figure}[!htb]
   \centering
\includegraphics[width=\textwidth]{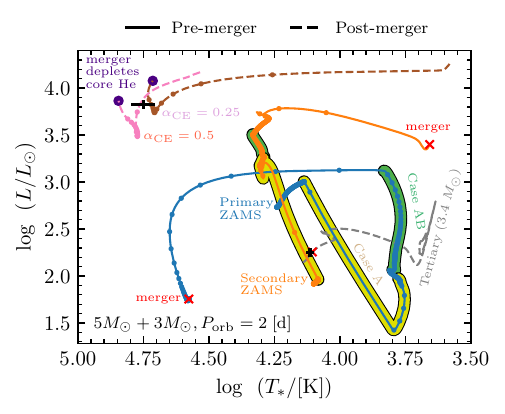}
    \caption{{\bf Models for our proposed evolutionary scenario for HD~45166.} \texttt{MESA} models are plotted on a $T_* - \log L$ diagram for a triple system comprising an inner binary with an initial period of 2\,d and initial masses of $5\,M_\odot$ (primary, blue line) and $3\,M_\odot$ (secondary, orange line), and a tertiary (gray line) with an initial mass of $3.4\,M_\odot$. Solid lines correspond to the evolution of the three components before the inner binary merges; dashed lines represent the post-merger evolution. Dots along the evolution tracks are separated by 0.1\,Myr of evolution. The merger position is marked with a red cross. Yellow and red highlighted regions correspond to mass transfer on the main sequence (case A) and after the main sequence  (case AB), respectively. The upper brown and pink tracks depict the evolution of the merger product (representing the Wolf-Rayet component) for common-envelope ejection efficiencies of $\alpha_{\rm CE} = 0.25$ and $0.5$, respectively \cite{MaterialsMethods}. Black plus signs mark the observed locations of the Wolf-Rayet and B7~V components. Purple filled circles mark the core He depletion of the merger, which occurs after 133.2 and 133.7\,Myr for $\alpha_\mathrm{CE}=0.25$ and $0.5$, respectively. 
    }
    \label{fig:hr_merger}
\end{figure}

We therefore propose that the Wolf-Rayet component in  formed from the coalescence of the helium cores of two intermediate-mass stars that were bound in a close binary. We construct an evolutionary model of this scenario  using the Modules for Experiments in Stellar Astrophysics (\texttt{MESA}) stellar evolution code \cite{Paxton2011} (Fig.\,\ref{fig:hr_merger}).  The system is started in a triple configuration with a tight inner binary and a distant tertiary representing the B7~V component. We find that the primary (more massive) star of the inner binary expanded and interacted with its companion, thereby losing its outer layers and becoming a stripped star, while the companion accreted material and became rejuvenated with hydrogen. 
% Such post-interaction binaries have been observed \cite{Gies1998, Shenar2020LB1, Bodensteiner2020}. 
Later, the secondary star expanded, leading to an unstable mass-transfer. This process leads to the formation of a gaseous envelope around the two stars, during which the two stars lose orbital energy due to friction with the envelope and spiral inwards (common-envelope evolution).  Due to the high binding energy of the hydrogen rich layers, this phase ends with both helium cores merging into a $\approx 2\,M_\odot$  magnetized helium star while most (but not all) of the hydrogen envelope is ejected. Being roughly 200 times more luminous than a main sequence $2\,M_\odot$ star \cite{Harmanec1988}, the merger product has a high luminosity-to-mass ratio. This, combined with its high effective temperature, launches a radiatively-driven outflow. In the absence of a magnetic field, such an outflow would not be easily detectable in a  spectrum. However, the trapping of the outflow by the magnetic field is known to increase the density of the circumstellar material and produce spectral emission lines like those observed in the Wolf-Rayet component \cite{Petit2013, Shenar2017Mag}. The distant tertiary star in our model does not impact the final outcome of the inner binary, except for perhaps catalyzing the merging of its stellar components \cite{Toonen2020}. Given its large orbital separation, we do not expect the tertiary star to show evidence for  accreted material from the merger ejecta.

% is the component currently seen as the secondary B7~V star, and its presence likely did not impact the final outcome of the inner binary, except for perhaps catalyzing the merging of its stellar components \cite{Toonen2020}. {\bf Given its large orbital separation, we do not expect it to show evidence for  accreted material from the merger ejecta.}

The scenario proposed here is quantitatively and qualitatively consistent with the observed properties of the system. The {\sc mesa} model reproduces the masses of the two components and the age of the system. The merger provides an explanation for the emergence of a magnetic field in the Wolf-Rayet component, and the magnetic field provides an explanation for the presence of emission lines in the spectrum of a 2\,$M_\odot$ helium star. 
% The scenario also provides an explanation for the presence of hydrogen, carbon, nitrogen, and oxygen in the spectrum of the Wolf-Rayet component \cite{Groh2008}, whose abundances depend  on the efficiency of material ejection ($\alpha_{\rm CE}$) during the common-envelope process \cite{MaterialsMethods}.
The binary nature of  enabled us to use the companion as a clock to constrain the evolutionary path of the system, and as a scale to determine the mass of the Wolf-Rayet component -- conditions that are rarely present in other proposed merger products \cite{Schneider2019, Gies2022}. 
Given the proximity of HD~45166 to Earth ($\approx 1\,$kiloparsec), other massive magnetic helium stars  have likely already been spectroscopically identified as Wolf-Rayet stars but not recognized to be magnetic (see Supplementary Text). 
% The most promising venue for finding them appears to be among Wolf-Rayet catalogs. Establishing a sample of super-Chandrasekhar mass magnetic helium stars will be key for studying their unique magnetospheres and their relative contribution to the formation of the magnetar population.

\bibliography{papers.bib}

\bibliographystyle{Science}

% \section*{Acknowledgments}

% {\bf Funding:} TS acknowledges support from the European Union's Horizon 2020 under the Marie Skłodowska-Curie grant agreement No 101024605. 
% DMB and TVR gratefully acknowledge support from Fonds Wetenschappelijk Onderzoek (FWO) by means of senior and junior postdoctoral fellowships under grant agreement Nos. 1286521N and 12ZB620N, respectively. This research has received funding from the European Research Council (ERC) under the European Union’s Horizon 2020 research and innovation programme (grant agreement number 772225: MULTIPLES).
% AG acknowledges support from a grant by the Prof.\ Amnon Pazy Research Foundation. The work of ANC is supported by NOIRLab, which is managed by the Association of Universities for Research in Astronomy (AURA) under a cooperative agreement with the National Science Foundation. NSL and GAW acknowledge financial support from the Natural Sciences and Engineering Research Council (NSERC) of Canada. ASO thanks FAPESP (grant 03/12618-7) for financial support.   \\

\noindent
{\bf Acknowledgments:} 
We thank the three anonymous referees for comments that improved our manuscript.
The {\scshape powr} code was developed under the guidance of Wolf-Rainer Hamann with substantial contributions from Lars Koesterke, Gotz Gr\"afener, Andreas Sander, Tomer Shenar and other co-workers and students. We thank Jason Hessels, Norbert Przybilla, and Huib Henrichs for helpful discussions.  The TESS, IUE, and FUSE data were obtained from the Mikulski Archive for Space Telescopes (MAST) at the Space Telescope Science Institute (STScI), which is operated by the Association of Universities for Research in Astronomy, Inc., under NASA contract NAS5-26555. Support to MAST for these data is provided by the NASA Office of Space Science via grant NAG5-7584 and by other grants and contracts. Funding for the TESS mission is provided by the NASA Explorer Program. TS, PM, LO, and HT thank the International Space Science Institute (ISSI, Bern) for hosting a discussion meeting. ANC was supported by NOIRLab, which is managed by the Association of Universities for Research in Astronomy (AURA) under a cooperative agreement with the National Science Foundation. This study is partly based on observations made with the Mercator Telescope, operated on the island of La Palma by the Flemish Community, at the Spanish Observatorio del Roquede los Muchachos of the Instituto de Astrofísica de Canarias. \\
{\bf Funding:} \\
TS was supported by the European Union's Horizon 2020 Marie Skłodowska-Curie grant No.\ 101024605. \\
DMB and TVR were supported by the Fonds Wetenschappelijk Onderzoek (FWO) by means of senior and junior postdoctoral fellowships under grant agreement numbers No.\ 1286521N and No.\ 12ZB620N, respectively.\\
HS was supported by the European Research Council (ERC) under the European Union’s Horizon 2020 research and innovation programme (No.\ 772225: MULTIPLES). \\
AG was supported by the Prof.\ Amnon Pazy Research Foundation.\\
NSL and GAW were received financial support from the Natural Sciences and Engineering Research Council (NSERC) of Canada.\\ 
ASO was supported by the FAPESP grant No.\ 03/12618-7.\\
ST was supported by the Netherlands Research Council NWO VENI 639.041.645 and VIDI 203.061.\\
{\bf Authors contributions:} TS developed the hypothesis, led the ESPaDOnS observations, performed the spectral and orbital analyses, and wrote the manuscript. GW performed the spectropolarimetric analysis and contributed to the observations and the manuscript. PM conceived the evolutionary scenario and constructed the corresponding \texttt{MESA} model. SB contributed to the analysis of spectropolarimetric data of HD~45166 and to the manuscript. JB collected HERMES data. DMB and TVR performed the light curve analysis. AG contributed to the evolutionary discussion and performed a population synthesis. NL, ST, and LO contributed to the discussion. ANC, NSL, and HS contributed to the  observations, orbital analysis, and the data reduction. LO contributed to the discussion on magnetic Wolf-Rayet stars. HS contributed to the design of the observational campaign and the orbital analysis. NSL contributed to the design of the observational campaign. ASO collected the LNA and FEROS observations. HT contributed to the spectral analysis aid in the spectral analysis. \\
{\bf Competing interests:} The authors declare no competing interests. \\
{\bf Data and materials availability:} The  
ESPaDOnS spectra are available on the CFHT archive (\url{www.cadc-ccda.hia-iha.nrc-cnrc.gc.ca/en/cfht}), proposal ID 22BC13, PI: Shenar. 
The FEROS data are available on the ESO archive (\url{archive.eso.org}) upon querying HD~45166 (no program ID available). The IUE {\bf and} FUSE data are available on the MAST archive (url{archive.stsci.edu})
% \cite{MASTARCHIVE} 
under program IDs WRJSH (PI: Heap),  JA017 (PI: Willis), and EI273 (PI: Stickland) for IUE, program ID P224 (PI: Willis) for FUSE.  The TESS FFI data used in this work to extract light curves are publicly available from the MAST archive with observation IDs tess-s0006-1-3 and tess-s0033-1-4 (PI: Ricker) for TESS,  and were downloaded and reduced interactively using the {\sc lightkurve} software \cite{lightkurve2018}. Reduced and wavelength-calibrated FEROS, ESPaDOnS, LNA and HERMES data and the input and output files of our stellar evolution {\sc mesa} model are available on Zenodo (\url{https://zenodo.org}) \cite{SpecEvZenodo}. The {\scshape powr} source code is available on \texttt{GitHub} (\url{https://github.com/powr-code/PoWR}) \cite{PoWRZenodo}. The derived RVs are provided as  supplementary materials (table S1) and are available on Zenodo \cite{SpecEvZenodo}.

% Any other processed materials  are available from the corresponding author upon reasonable request.

%Here you should list the contents of your Supplementary Materials -- below is an example. 
%You should include a list of Supplementary figures, Tables, and any references that appear only in the SM. 
%Note that the reference numbering continues from the main text to the SM.
% In the example below, Refs. 4-10 were cited only in the SM.     
% \section*{Supplementary materials}

\section*{List of Supplementary Materials}

Materials and Methods\\
Supplementary Text\\
Figs. S1 to S11\\
Tables S1 to S2\\
References \textit{(51-110)} \\
table S1 

\clearpage

% \section*{Tables}

% \section*{Figures}

% \begin{figure}[!h]
%   \centering
% \includegraphics[width=\textwidth]{Figs/Sketch_qWR.pdf}
%     \caption{{\bf Visualisation of HD~45166.} This binary is found here to comprise a $3.4\,M_\odot$ B7~V main-sequence star and  a hot ($T_{\rm eff} = 56\,$kK) $2\,M_\odot$ Wolf-Rayet star that possesses the strongest magnetic field ever measured for a non-degenerate super-Chandrasekhar mass object. The orbital period is found to be $22.5\,$yr and the orbital separation 10.5\,au. Credit: F.\ Bodensteiner; background: ESO/B.\ Tafreshi (twanight.org).} 
%     \label{fig:Visu}
% \end{figure}

\clearpage

\newpage

% \section*{Supplementary Materials}

\section*{Materials and methods}

% \counterwithin{figure}{section}

\newcommand{\hbAppendixPrefix}{S}
\renewcommand{\thefigure}{\hbAppendixPrefix\arabic{figure}}
\setcounter{figure}{0}
\renewcommand{\thetable}{\hbAppendixPrefix\arabic{table}} 
\setcounter{table}{0}
\renewcommand{\theequation}{\hbAppendixPrefix\arabic{equation}} 
\setcounter{equation}{0}

\subsection*{Observational data} 

We acquired spectropolarimetric observations (Stokes $I$ and $V$) of HD~45166 on 2022 February 17 and 19 with the ESPaDOnS spectropolarimeter \cite{Donati2006ESP} mounted on the CFHT (program ID 22AC08; PI: Shenar). A total of eight 1-hr exposures (four per night), each divided into four sub-exposures, resulted in typical signal-to-noise (S/N) of $\approx 150-200$ per wavelength bin at $\approx 5000\,$\AA\  and a resolving power $R = 65\,000$. The spectra cover the wavelength range 3668 --  10\,480\,\AA. The data were reduced and wavelength-calibrated using CFHT's standard {\scshape upena} pipeline \cite{Martioli2011}. Including the subexposures, a total of 32 Stokes $I$ and $V$ ESPaDOnS spectra were obtained. The co-added ESPaDOnS $I$ and $V$ spectra are shown in Fig.\,\ref{fig:Zeeman}.

Additionally, we measured RVs from archival optical spectra obtained with three additional spectrographs. The first set of archival data  was acquired between 1998 to 2004  using the Coud\'e spectrograph at the
1.6\,m telescope of the LNA
in Itajub\'a, Brazil \cite{CoudeLNAManual}, and has been previously described by \cite{Steiner2005}. The majority of LNA spectra cover a modest spectral range (4520 -- 4960\,\AA) at a resolving power of $R \approx 7000$, reaching a typical S/N of 20 -- 30 in the continuum. A few additional higher-resolution spectra ($R \approx 20\,000$) covering the range 3830 -- 4580\,\AA\ were also used. In total, 103 LNA spectra were available. 

The second set of archival spectroscopic data  were obtained in 2002 with the FEROS instrument \cite{Kaufer1999} mounted at the 1.52\,m telescope of the European Southern Observatory (ESO) in La Silla, Chile \cite{Kaufer1999}. These data cover a spectral range 3830 -- 9215\,\AA\ at a resolving power of $R = 48\,000$, with typical S/N of 80 in the range 4500 -- 7000\,\AA\, decreasing down to $\approx 35$ at the blue and red edges. A total of 36 FEROS spectra are available, taken at a high cadence over a few nights. The data reduction and calibration were also described previously \cite{Steiner2005}.

The third and final archival spectroscopic dataset was acquired in 2019 -- 2021, with the exception of a single spectrum in 2012, using the {\scshape HERMES} spectrograph \cite{Raskin2011} mounted on the 1.2\,m Mercator telescope at the Observatorio del Roque de Los Muchachos on La Palma, Spain \cite{Raskin2011}. HERMES spectra cover the wavelength range from 3770 to 9000\,\AA\ with a spectral resolving power of $R\approx$\,85\,000. A total of 28 HERMES spectra were available, having a typical S/N of 30 in the continuum. Standard reductions including bias and flat-field corrections and wavelength calibrations were performed using the HERMES data reduction pipeline \cite{Raskin2011}. Barycentric corrections were applied.

All spectra were normalized by fitting a piecewise linear function through identified continuum regions selected homogeneously for all datasets. 
Given the low-amplitude motion of the binary components, stable wavelength calibration is essential.  Relative wavelength calibration was brought to a precision  of $\approx 0.1\,\kms$ by cross-correlating the interstellar sodium lines Na\,{\scshape i}\,$\lambda \lambda 5890, 5896$ in the FEROS, HERMES, and ESPaDOnS spectra with a selected ESPaDOnS spectrum (which is wavelength-calibrated). However, this was not possible for the LNA spectra, which do not cover these lines. We therefore consider the RVs of the LNA data as less reliable, and allow for a systematic RV shift between them and the other datasets (see below). 

For the spectral analysis, we append the optical spectra with a set of UV spectra retrieved from the Mikulski Archive for Space Telescopes (MAST). The data were acquired with the International Ultraviolet
Explorer (IUE) (program IDs: WRJSH, JA017; PIs: Heap, Willis, respectively) and presented previously \cite{Willis1983, Willis1989}. The spectra cover  1150 -- 2150\,\AA\ at a resolving power of $R\approx 10\,000$. There are substantial changes in P-Cygni lines that occur over short and long timescales \cite{Willis1983, Willis1989}.  We use a co-added spectrum formed from 20  high-cadence IUE spectra obtained consecutively on 1988 February 3, 4, and 5  for 36\,hr.  The individual spectra have a S/N of 5-6, such that the co-added spectrum has a total S/N of $\approx 20$. While the process of co-adding smears the small-amplitude variability of the Wolf-Rayet component on the 5\,hr period (see below), the impact is negligible compared to the spectral resolution. Since we only use this spectrum to estimate the effective temperature of the Wolf-Rayet component and analyze the SED, the process of co-adding the data has no impact on our results. The spectrum is normalized using the model continuum obtained from the SED analysis (see below).

For the analysis of the SED, we also retrieved from MAST a far-UV spectrum covering the spectral range  900 -- 1200\,\AA\ obtained with the  Far Ultraviolet Spectroscopic Explorer (FUSE), which have also been previously described  \cite{Willis2006}. The spectrum was acquired on 2001 Mar 5 (PI: Willis, program ID: P224), 
 has  a resolving power of $R \approx 20\,000$, and a S/N of $\approx 100$. Since the FUSE and IUE spectra overlap, we use the IUE spectra to calibrate the FUSE spectrum rather than the standard flux calibration, such that the overlapping regions ($\approx$ 1150 -- 1200\,\AA) of the IUE spectra and FUSE spectra agree, resulting in a multiplicative factor of 0.8 of the FUSE spectrum. For the SED analysis, we append the FUSE and IUE spectra with a low-resolution flux calibrated IUE spectrum obtained on 1982 Sep 06 (PI: Stickland, ID: EI273), which covers the range 1850 -- 3350\,\AA\ at a resolving power of $R\approx 500$ and S/N$\approx 5$.

Finally, for the SED analysis, we use $U$ photometry \cite{Abazajian2009}, $BV$ photometry \cite{Zacharias2004}, $JHK$ photometry \cite{Bonanos2009}, and Wide-field Infrared Survey Explorer (WISE) photometry  \cite{Wright2010}. 

\subsection*{Analysis of the spectropolarimetric data}

% As described in the main part of the paper, high resolution circular polarization (Stokes $V$ and $I$) spectra of  were obtained on 17/19 February 2022 using the ESPaDOnS spectropolarimeter at the CFHT. Four 2-hour observations were obtained: two on February 17, and two on February 19. Clear Stokes $V$ variations were immediately apparent across many spectral lines corresponding to the WR component. Following examination of the individual observations it was concluded that there was no significant variability detected in the WR spectral lines or the Stokes $V$ signatures, and the spectra were co-added to achieve a higher S/N. 

% As described above, the ESPaDOnS spectra  Clear Stokes $V$ variations were immediately apparent across many spectral lines corresponding to the WR component. Following examination of the individual observations it was concluded that there was no significant variability detected in the WR spectral lines or the Stokes $V$ signatures, and the spectra were co-added to achieve a higher S/N. 

% \noindent{\bf General evidence for magnetism}

Stokes $V$ variations are detected in all ESPaDOnS spectra
% across essentially all spectral lines attributed to the Wolf-Rayet component 
in multiple  lines attributed to the Wolf-Rayet component (Fig.\,\ref{fig:Zeeman}), most prominently lines of He~{\scshape ii} and lines of highly ionized elements. 
Most lines are in emission in the intensity spectrum, although O\,{\scshape v}\,$\lambda 5114$ appears in absorption. No similar Stokes $V$ features are observed in the narrow absorption lines belonging to the B7~V component (Fig.~\ref{fig:Zeeman}I). 
Since the four individual ESPaDOnS Stokes $V$  spectra obtained during the two observing nights do not show notable variability in the context of the magnetic properties, they were co-added to achieve a higher S/N.

The circular polarization features are qualitatively consistent with the Zeeman effect due to the presence of a strong magnetic field in HD~45166. While the features exhibit some morphological diversity (Fig.~\ref{fig:Zeeman}), they are all - except for O\,{\scshape v}\,$\lambda 4930$ - of the same polarity, with positive circular polarization observed in the blue wing, and negative polarization observed in the red wing.  O\,{\scshape v}\,$\lambda 4930$ is a weak emission line that appears distinct from the other emission features, with  a polarity opposite to that of all other lines in the spectrum (Fig.~\ref{fig:Zeeman}D). 

The O\,{\scshape v}\,$\lambda 5114$ and $4930$ lines indicate the magnetic nature of the Wolf-Rayet component and provide quantitative information about its magnetic field. Both lines exhibit resolved Zeeman splitting into triplets that correspond in wavelength to the locations of the extrema and inflection points of their associated Stokes $V$ features. Their narrow profiles, and the pure absorption profile of the O\,{\scshape v}\,$\lambda 5114$ line, indicate that they form in or close to the stellar surface. 
The inverted Stokes $V$ signature of O\,{\scshape v}\,$\lambda 4930$ can be  explained by the Zeeman effect operating on a spectral line formed in emission, and not in absorption. As discussed previously \cite{Wade2012} for the strongly magnetic O-type star NGC~1624-2, spectral lines of hot stars can form in emission under non-LTE conditions even in or close to the stellar surface.

To quantify the strength and geometry of the detected magnetic field, we measured the mean magnetic field modulus $\langle B\rangle$ and mean longitudinal magnetic field $\langle B_{\rm z}\rangle$  from the O\,{\scshape v}\,$\lambda\lambda 5114$ and $4930$ lines. We renormalized the two lines to the local continuum.  We measured the mean field modulus by fitting Gaussians to the three components of each Zeeman triplet to measure the average separation between the two polarized line components and the unpolarized line, $\Delta\lambda_{\rm Zeeman}$. We then compute  $\langle B\rangle=\Delta\lambda_{\rm Zeeman}/(4.67 \times 10^{-13}\,g_{\rm l}\,\lambda_0^2)$, where $g_{\rm l}=1.0$ is the theoretical Land\'e factor of the line and $\lambda_0=4930$\,\AA\ or $5114$\,\AA\ is the rest wavelength.   We find
$\langle B\rangle_{\rm qWR}=44.1\pm 2.0$\,kG and  $\langle B\rangle_{\rm qWR}=41.8\pm 1.0$~kG for the O\,{\scshape v}\,$\lambda\,4930$ and $\lambda 5114$ lines, respectively.  The shallow depths of the two O~{\scshape v} lines and uncertainty in the positioning of the continuum leads to  systematic errors that are not included in the formal $1\sigma$ unceratinties provided, which we thus ignored when computing the mean and standard deviation: $\langle B\rangle_{\rm qWR}=43\pm 2.5$\,kG.  We measured a mean longitudinal field of  $\langle B_{\rm z}\rangle_{\rm qWR}=16.1\pm 2.0$\,kG and $\langle B_{\rm z}\rangle_{\rm qWR}=11.1\pm 1.0$\,kG for the O\,{\scshape v}\,$\lambda 4930$ and $\lambda 5114$ lines, respectively, using the first-order moment method \cite{wade2000},  adopting the same Land\'e factor and rest wavelengths. Taking the mean and standard deviation yields  $\langle B_{\rm z}\rangle_{\rm qWR}=13.5\pm 2.5$\,kG. As before, we do not weigh the values by the errors due to unaccounted systematic errors.

\begin{figure}[!htb]
   \centering
\includegraphics[width=0.6\textwidth]{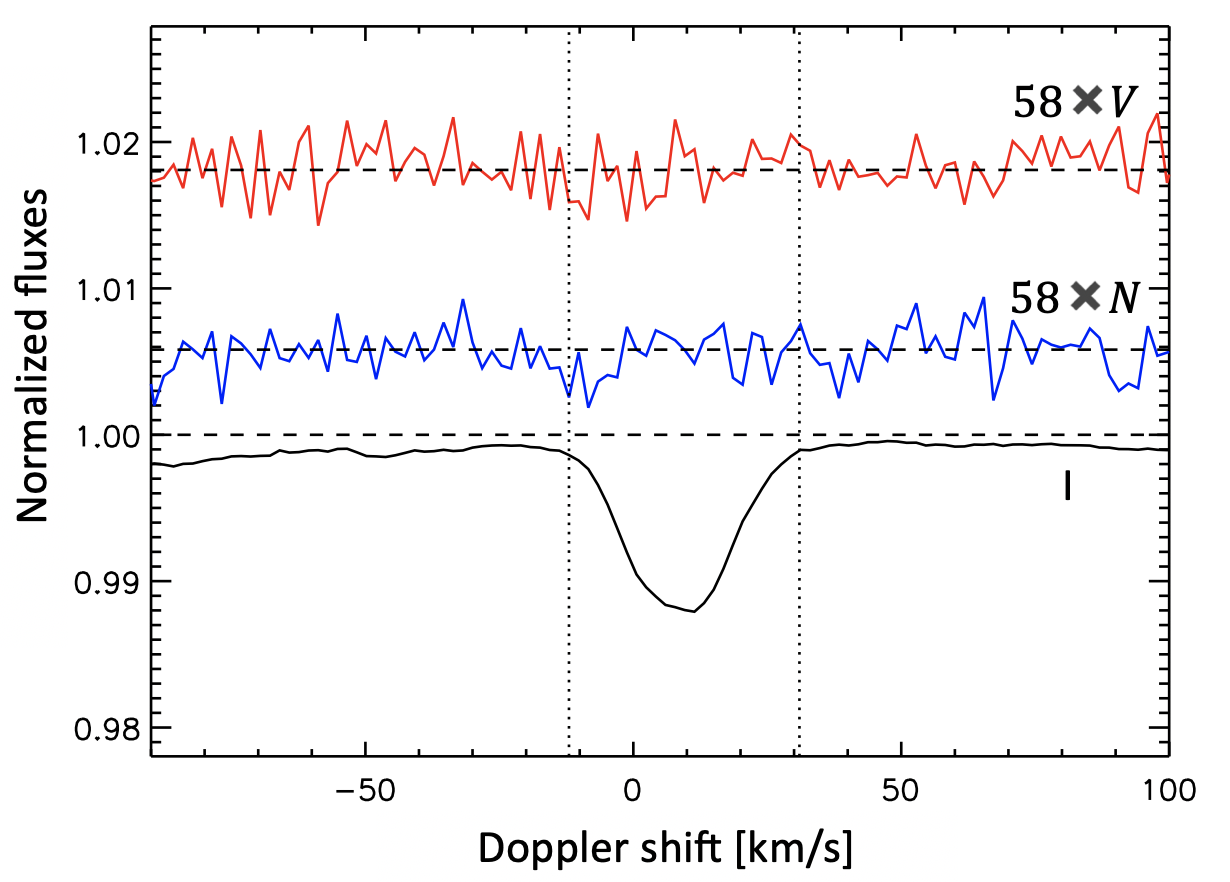}
    \caption{{\bf LSD profile of the B7~V companion.} The lower (black) curve is the mean Stokes $I$ profile. The upper (red) curve is the measured net circular polarization (Stokes $V$). The middle (blue) curve is the diagnostic null $N$. No signal is detected inside or outside the line range in either $V$ or $N$. The vertical dashed lines indicate the integration range used for measurement of the detection probability and longitudinal magnetic field.
    } 
    \label{fig:B7lsd}
\end{figure}

 We also extracted Least-Squares Deconvolution (LSD) mean profiles of the B7V companion (Fig.\,\ref{fig:B7lsd}). We employed a standard solar abundance B7 line mask \cite{Wade2016} and used the {\scshape iLSD} code \cite{kochukhov2010, iLSDLink}. From the LSD profile we determined the detection probability in both Stokes $V$ and the diagnostic null (both non-detections), and determined the upper limit of the star's mean longitudinal magnetic field [$< 60$\,G at $3\sigma$ confidence; \cite{wade2000}].

\begin{figure}[!htb]
   \centering
   \includegraphics[width=\textwidth]{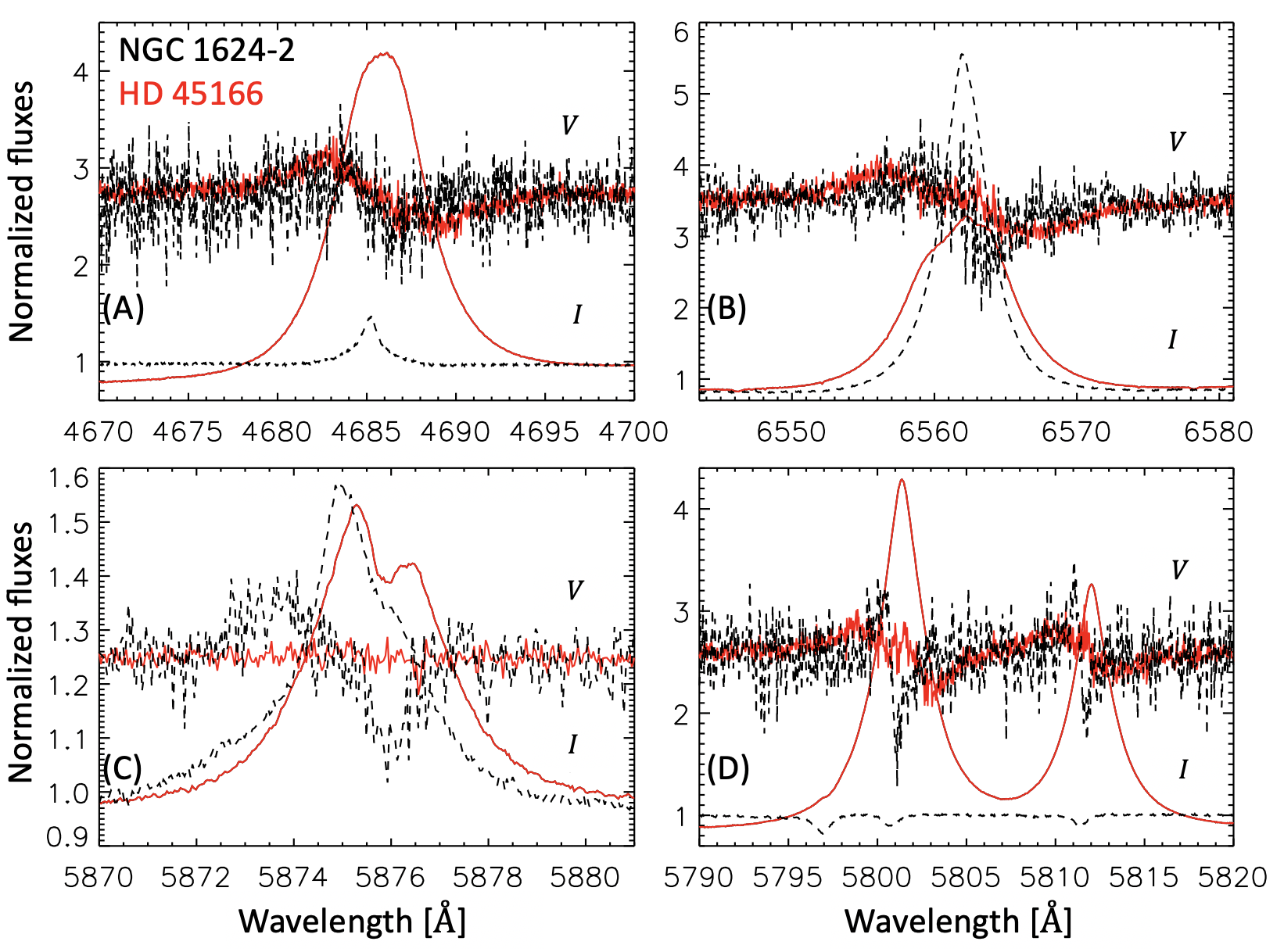}
% \vspace{0.5cm}
    \caption{{\bf Comparison between HD~45166 and NGC~1624-2.} Shown are Stokes $I$ and $V$ line profiles (see labels in panels) in spectra of HD~45166 (red/solid) and NGC~1624-2 (black/dashed) for He~{\scshape ii}\,$\lambda$\,4686 (A),  H$\alpha$ +He\,{\scshape ii}\,$\lambda 6560$ (B), He~{\scshape i}\,$\lambda$\,5876 (C), and C~{\scshape iv}\,$\lambda\lambda$\,5801,5811 (D).
    } 
    \label{fig:NGCcomp}
\end{figure}

We perform a more general comparison between the Stokes $I$ and $V$ spectra of the Wolf-Rayet component and NGC~1624-2 (Fig.~\ref{fig:NGCcomp}). NGC~1624-2 is a magnetic star showing the Of?p phenomenon \cite{Walborn1972, Naze2010} with a very strong magnetic field of $\approx 20\,$kG \cite{Wade2012}, sufficiently large to produce marginal Zeeman splitting of some Stokes $I$ and Stokes $V$ profiles of some individual spectral lines. The star is known to have a large, dense magnetosphere \cite{Petit2015, DavidUraz2019}. The lower effective temperature of NGC~1624-2  of $35\pm 2$~kK \cite{Wade2012} leads to absorption lines of He~{\scshape i} in its spectrum. As a consequence, essentially all H and He~{\scshape i} lines, whether in emission or absorption,  show Zeeman signatures. In  HD~45166, however,
% while it shows Zeeman signatures associated with H lines, 
no Zeeman signature associated with  He~{\scshape i} lines is seen, even though  He~{\scshape i} lines are present (in emission) in its spectrum (Fig.\,\ref{fig:Zeeman}H). Instead, Zeeman signatures are detected in association with numerous lines of highly ionized light elements, in particular C~{\scshape iv}, N~{\scshape iv}, N\,{\scshape v}, and O~{\scshape v}. Another difference  is in the ratio of the He\,{\scshape ii}\,$\lambda 4686$ line to the H$\alpha$ + He\,{\scshape ii}\,$\lambda 6560$ blend. While the H$\alpha$ + He\,{\scshape ii}\,$\lambda 6560$ blend is of similar strength in both stars, the He\,{\scshape ii}\,$\lambda 4686$ line is almost 20 times stronger in the Wolf-Rayet component of HD~45166,  indicating that it is helium rich.

\begin{figure}[!htb]
   \centering
\includegraphics[width=0.5\textwidth]{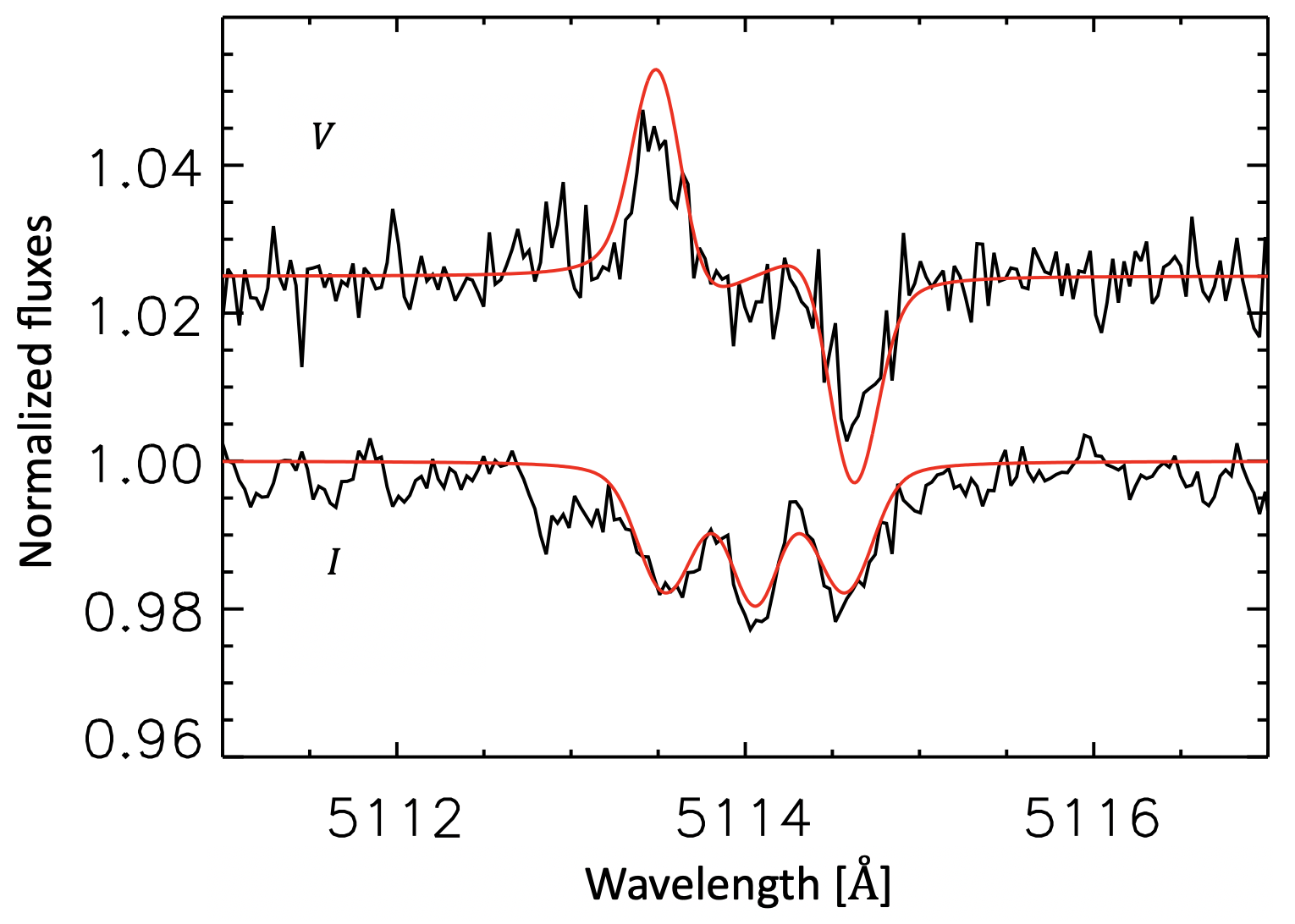}
    \caption{{\bf Modeling of the spectropolarimetric signature.} Shown are Stokes $I$ and $V$ profiles of O~{\scshape v}\,$\lambda 5114$ (black) compared to LTE polarized spectrum synthesis calculations (red) computed assuming a pole-on magnetic dipole of polar strength 45~kG. Such a field geometry roughly reproduces the observed profiles. 
    } 
    \label{fig:Z21fit}
\end{figure}

The observed properties of the magnetic field in the Wolf-Rayet component are similar to those found in other hot magnetic stars with dipolar magnetic fields \cite{Petit2013}, indicating 
 that the Wolf-Rayet component of  also hosts a global  magnetic field. In this configuration, the polarized flux is largely generated near the stellar surface, where the magnetic field is strongest. As in other magnetic star, the emission lines then originate from stellar wind trapped by the closed magnetic loops (the magnetosphere).  The Stokes-$V$ Zeeman signature is produced predominently in the surface layers of the star. This is because the Zeeman signature is seen only in high ionization lines, which are formed in or close to the stellar surface according to the model atmospheres computed for the star (see below). At larger radial extent, no or weak Zeeman signature is seen, both because the magnetic field strength rapidly drops radially ($\propto r^{-3}$ in a dipolar field, $r$ being the radial distance), and because the signature is diluted in the larger emitting volume, which exhibits opposing polarities. 
While some contribution to the observed circular polarization could be produced by field lines extending into the star's wind, the lack of any circular polarization detected in the emission lines of He~{\scshape i} indicates that such contributions are very small.  In Fig.~\ref{fig:Z21fit} we show LTE polarized spectrum synthesis calculations with the {\scshape zeeman} code \cite{landstreet1988,wade2001, SpecEvZenodo} for the O~{\scshape v}\,$\lambda 5114$ line, adopting a dipolar magnetic field of 45~kG  with the magnetic field axis parallel to Earth’s line-of-sight. The narrow Zeeman components constrain the non-thermal line broadening to $\lesssim$10\,km/s. In Fig.~\ref{fig:Z21fit} we have adopted a Doppler width of 6~km/s to account for rotational and turbulence broadening. Although we have adopted a simplified magnetic geometry, the Stokes $I$ and $V$ profiles are reproduced by this basic model.

The Stokes $V$ spectrum thus indicates that the underlying intensity spectrum of the Wolf-Rayet component, which is largely hidden behind the magnetosphere, is characterized by the presence of lines belonging to highly ionized species such as N\,{\scshape v}, O\,{\scshape v} and the lack of lower-ionization lines such as He\,{\scshape i} and O\,{\scshape iii}. To fulfill such conditions, the Wolf-Rayet component needs to be hot ($T_* \gtrsim 45\,$kK),  consistent with our spectral analysis (see below).

\subsection*{Spectral analysis}

To derive the stellar parameters of the components, we use the non-LTE Potsdam Wolf-Rayet ({\scshape powr}) model atmosphere code \cite{Graefener2002, Hamann2003, Sander2015, PoWRZenodo}, which performs radiative transfer in an expanding atmosphere. Models are defined by the chemical abundances, the surface effective temperature $T_*$, the bolometric luminosity $L$, the surface gravity $g$, and wind parameters such as the mass-loss rate $\dot{M}$ and terminal wind speed $v_\infty$. The radius $R_*$ is given by the Stefan-Boltzmann relation, $R_* \propto \sqrt{L}\,T_{\rm *}^{-2}$.  In {\scshape powr}, the effective temperature is  defined with respect to a continuum Rosseland optical depth of $\tau_{\rm Ross} = 20$ as a proxy for the hydrostatic layers. These layers are deeper than the photosphere ($\tau_{\rm Ross} = 2/3$), at which the photospheric effective temperature $T_{\rm eff}$ is defined.  For a star with a negligible wind (such as the B7~V component), $T_* \approx T_{\rm eff}$. For the Wolf-Rayet component, however, $T_{\rm eff}$ is the effective temperature with respect to a layer that is embedded in the circumstellar material. Hence, we use $T_*$ when comparing to evolution models.

In the subsonic regime, the density is determined by hydrostatic equilibrium. In the supersonic regime, the radial velocity field is assumed to follow a $\beta$-law \cite{Castor1975}, and the  density follows from the continuity equation. The supersonic domain does not impact our analysis, since the wind of the B7~V component is negligible, and since the emission features of the Wolf-Rayet component are ignored. In the main non-LTE iteration, the line profiles are assumed to have a constant Doppler width of  $30\,\kms$. During the integration, the Doppler widths are determined from the depth-dependent thermal broadening and microturbulence $\xi$ \cite{Shenar2015}.

\subsubsection*{Analysis of the B7~V component} 

% \subsection*{Spectral analysis, and the mass of the B7~V component} 

\begin{figure}[!htb]
   \centering
\includegraphics[width=\textwidth]{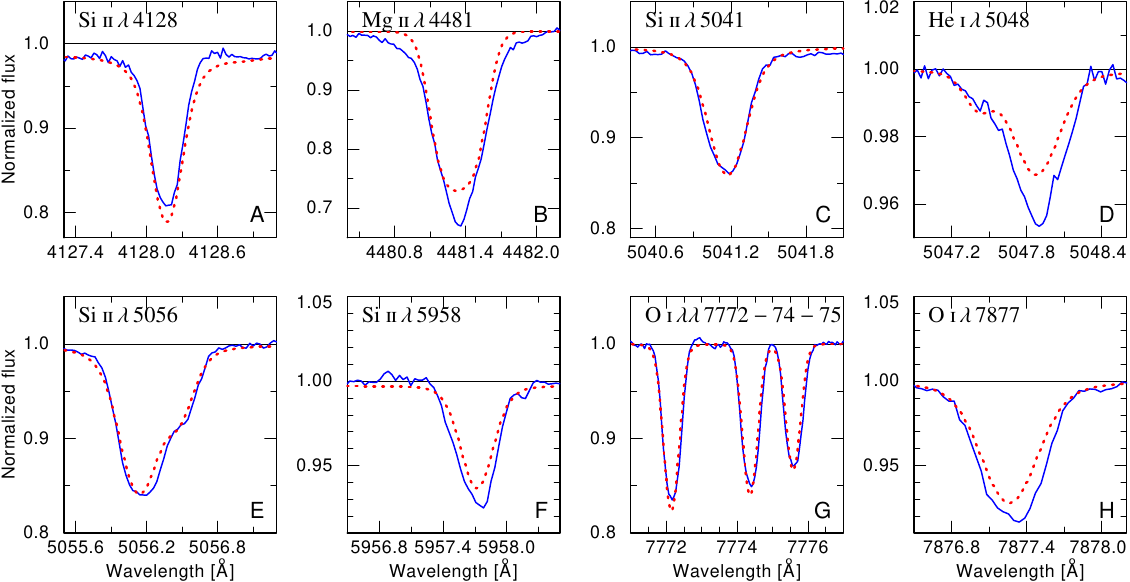}
    \caption{{\bf Spectral analysis of the B7~V component.} Shown is a comparison between our composite spectral model of HD~45166 (red) and the co-added ESPaDOnS spectrum (blue), focusing on absorption lines belonging to the B-type component. The composite model includes both the Wolf-Rayet and B7~V components, weighed with their respective wavelength-dependent light ratios. The spectral lines shown are formed only in the B7~V component, with the possible exception of He\,{\scshape i}\,$\lambda 5048$ (D), which may be contaminated by the magnetosphere of the Wolf-Rayet component.  Horizontal black lines are continuum levels. 
    } 
    \label{fig:Blines}
\end{figure}

Atmospheric abundances of the B7~V component are fixed to solar \cite{Asplund2009}, which is consistent with the data. The surface gravity of the B7~V component cannot be determined, since the diagnostic Balmer lines are entangled with magnetospheric emission and absorption originating in the Wolf-Rayet component in a non-trivial manner. We therefore assume $\log g / \cms = 4.0$ \cite{Harmanec1988}; varying  $\log g$ by $0.2\,$dex has little impact on the remaining diagnostic lines. Wind parameters are set to negligible values given the lack of wind diagnostics. Microturbulence is set to $\xi = 5\,\kms$.   We determine $T_*$ and the overall light contribution of the B7~V component in the $V$-band, $l_{\rm B}(V)$, using  diagnostic lines of O\,{\scshape i}, Mg\,{\scshape ii}, and Si\,{\scshape ii}, avoiding He\,{\scshape i} due to possible cross-contamination by the magnetosphere of the Wolf-Rayet component (Fig.\,\ref{fig:Blines}).  The projected rotational velocity $v \sin i$ and macroturbulent velocity $v_{\rm mac}$ are derived using a Fourier method \cite{Gray1992}, implemented by the  {\scshape iacob-broad}  tool \cite{Simon-Diaz2007, Simon-Diaz2014, IACOBURL}. However, in the domain where $v \sin i$ is small, this method is known to lead to degeneracies \cite{Simon-Diaz2014}, and we therefore only provide upper limits in Table\,\ref{tab:Parameters}. The rotation and macroturbulence are accounted for by convolving the synthetic spectrum of the B7~V component with corresponding rotational profiles with the value corresponding to the upper limit given in Table\,\ref{tab:Parameters}.   the projected rotational velocity of the B7~V component ($v \sin i \lesssim 10\,\kms$) is low for this spectral type \cite{Abt2002}, though other comparably slow rotators exist. While a pole-on configuration of the rotational axis of the B7~V component would provide an explanation for this, such a configuration would not be consistent with the orbital inclination derived below ($i=49\pm11^\circ$), unless it the orbital and rotational axes are misaligned.   More likely, the B7~V star is intrinsically a slow rotator.

\subsubsection*{Analysis of the Wolf-Rayet component} 

\begin{figure}[!htb]
   \centering
\includegraphics[width=.8\textwidth]{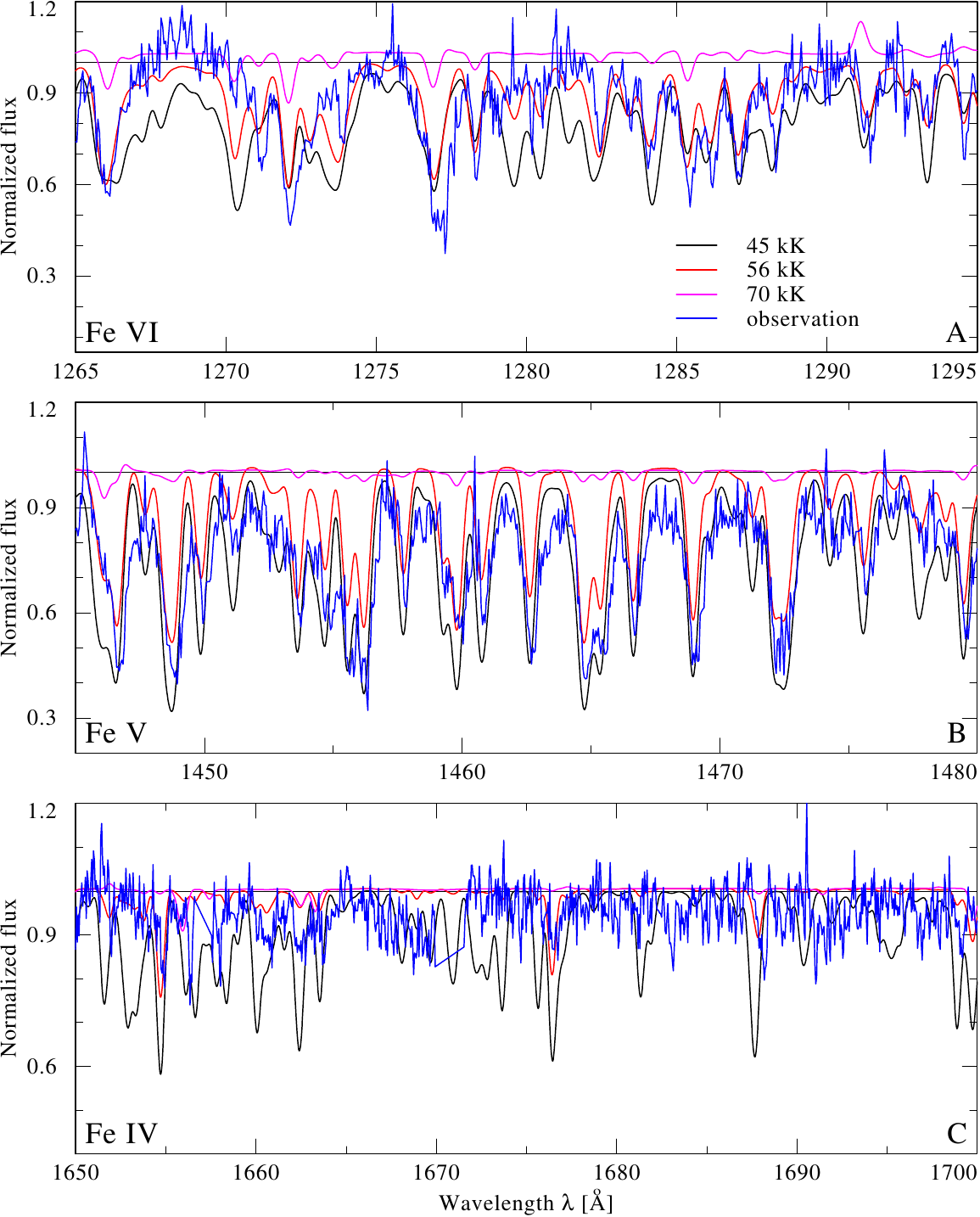}
    \caption{{\bf Spectral analysis of the Wolf-Rayet component}. Shown is a comparison between the co-added IUE spectrum (blue), normalized using the model continua, and three composite {\scshape powr} models calculated with $T_* = 45, 56$, and $70\,$kK (in black, red, and green, respectively) for the Wolf-Rayet component. The panels show regions dominated by Fe\,{\scshape vi, v}, and {\scshape iv} (panels A, B, and C, respectively). The contribution of the B7~V component is negligible (see Fig.\,\ref{fig:SED}). The effective temperature of the Wolf-Rayet component is estimated at $T_* = 56.0 \pm 5.0\,$kK. 
    } 
    \label{fig:IronForest}
\end{figure}

For the Wolf-Rayet component, we adopt previously derived elemental abundances for He, C, N, and O \cite{Groh2008}; $X_{\rm He} = 0.67$, $X_{\rm C} = 0.0059$, $X_{\rm N} = 0.002$, and $X_{\rm O} = 0.0015$, which leaves $X_{\rm H} = 0.32$ (values are mass fractions of the respective element). The remaining abundances are set to solar.  Implications of this abundance pattern have been discussed previously \cite{Groh2008}; it is anomalous compared to either Wolf-Rayet stars or central stars of planetary nebulae. However, these abundances were derived assuming that the emission features stem from a radially expanding outflow, disregarding the magnetic confinement of the wind.  We therefore refrain from interpreting the abundances quantitatively. Unlike the previous investigation  \cite{Groh2008},  we rely solely on the iron forest in the UV, which form in the stellar surface and are not contaminated by the magnetosphere.  For this purpose, the co-added IUE spectrum is used, which is normalized using the added model continua of both stars. We adopt a mass-loss rate of \mbox{$\log \dot{M}/$(\smy) = $-9.0\,$}~for the Wolf-Rayet component; The high mass-loss rate of \mbox{$\log \dot{M}$/(\smy) = $-6.7\,$}~previously derived \cite{Groh2008} was a consequence of ignoring the magnetosphere, so we do not adopt it.  We use regions dominated by lines belonging to Fe\,{\scshape iv, v, vi} (Fig.\,\ref{fig:IronForest}).  By using the UV lines, the contribution of the B-type companion  is negligible  (see SED analysis below). The narrow absorption features in the visual imply a combined projected rotational velocity and microturbulence $v \sin i, \xi \lesssim  10\,\kms$, and we adopt these upper limits for the computation of the spectrum. 
The low projected rotational velocity of the Wolf-Rayet component can be understood as a consequence of angular momentum loss via the magnetic field \cite{Weber1967, Ud-Doula2009}. From the iron forest, we determine $T_* = 56\pm5\,$kK, which is lower than the value of 70\,kK derived previously \cite{Groh2008} by analyzing magnetospheric features using a wind model.

\subsubsection*{Spectral energy distribution and age} 

\begin{figure}[!htb]
   \centering
\includegraphics[width=\textwidth]{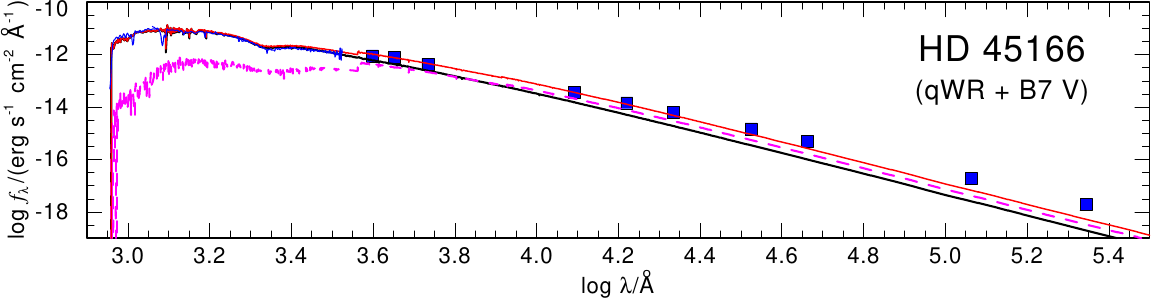}
    \caption{{\bf Spectral energy distribution of HD~45166.} The observed SED (blue line for IUE and FUSE spectroscopy and blue squares for photometry) is compared to the sum (solid red line) of the synthetic SEDs of our models for the Wolf-Rayet component (solid black line) and the B-type component (dashed green line). The Wolf-Rayet component fully dominates the UV flux. The flux excess observed in the infrared could originate in free-free emission stemming from the trapped outflow.  } 
    \label{fig:SED}
\end{figure}

Using the {\scshape powr} models of the two components and their light ratio in the visual, we compute the luminosities and reddening by fitting the multiwavelength SED to the sum of the models (Fig.\,\ref{fig:SED}). We use a reddening law \cite{Cardelli1989} with a total-to-selective extinction ratio of $R_V = 3.1$. We find $\log L_{\rm qWR}/L_\odot = 3.830\pm0.050$ and $\log L_{B}/L_\odot = 2.250\pm0.050\,[L_\odot]$, with a color excess of $E_{B - V}  = 0.210\pm0.010$, which is consistent with the strength of the interstellar lines (e.g., Ly$\alpha$). The parameters derived for the B-type component are  consistent with its spectral type \cite{Harmanec1988}.  A flux excess is noted in the infrared which grows gradually between $2\,{\mu}$m and $22\,{\mu}$m; it could originate in free-free emission stemming from the trapped outflow, which is not included in our model.

We used the {\scshape BONNSAI} Bayesian tool \cite{Schneider2014, BONNSAI} to estimate the current mass and age of the B-type component using single-star evolution tracks \cite{Brott2011}, using our derived $T_*$, $\log L$, and $v \sin i$ as input parameters. We find a mass of $M_{\rm B} = 3.38\pm0.10\,M_\odot$ and an age of $105\pm 35\,$Myr, which we consider as the age of the  binary.

\subsection*{TESS light curve and the 1.6\,d period}

\begin{figure}[!htb]
   \centering
\includegraphics[width=0.46\textwidth]{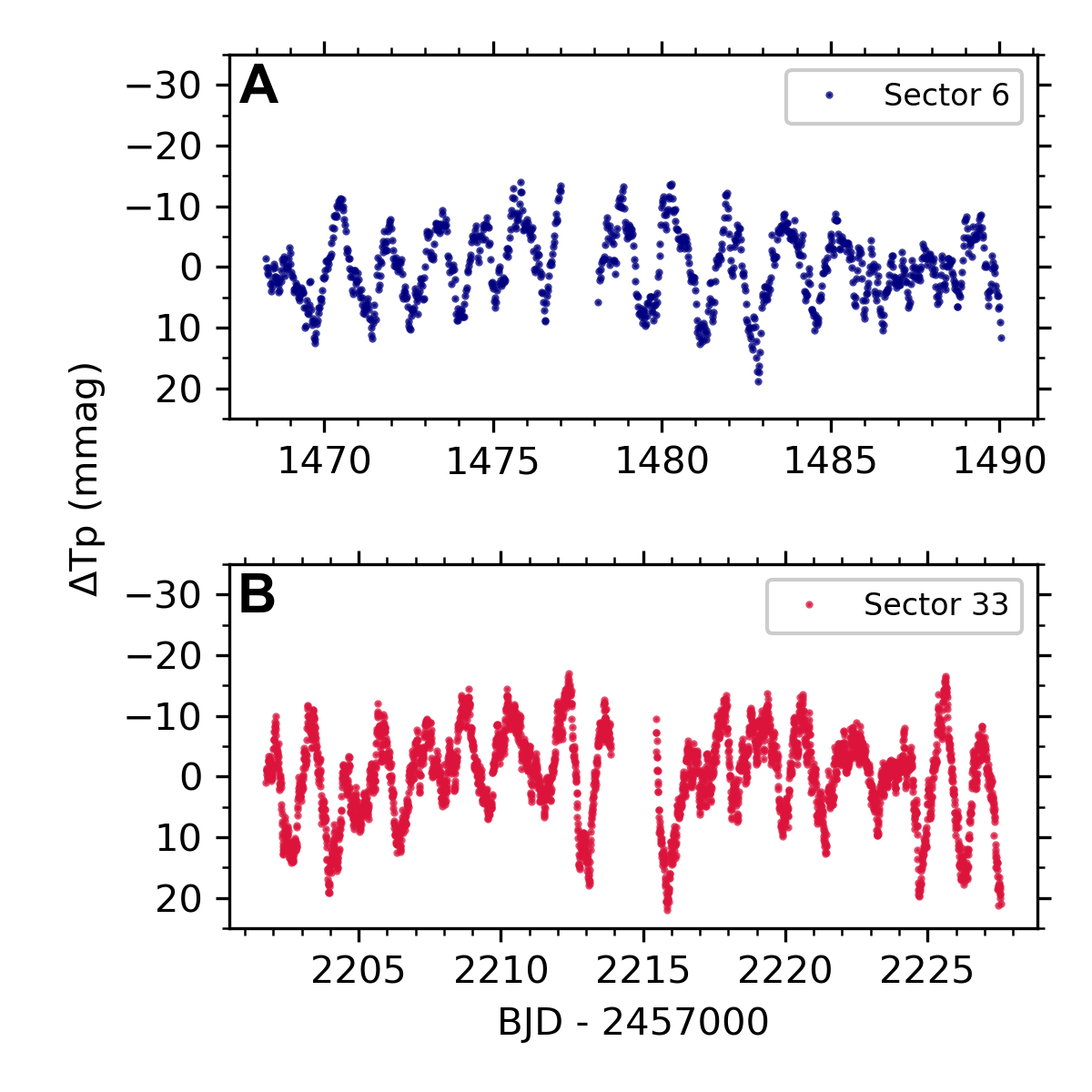}
\includegraphics[width=0.46\textwidth]{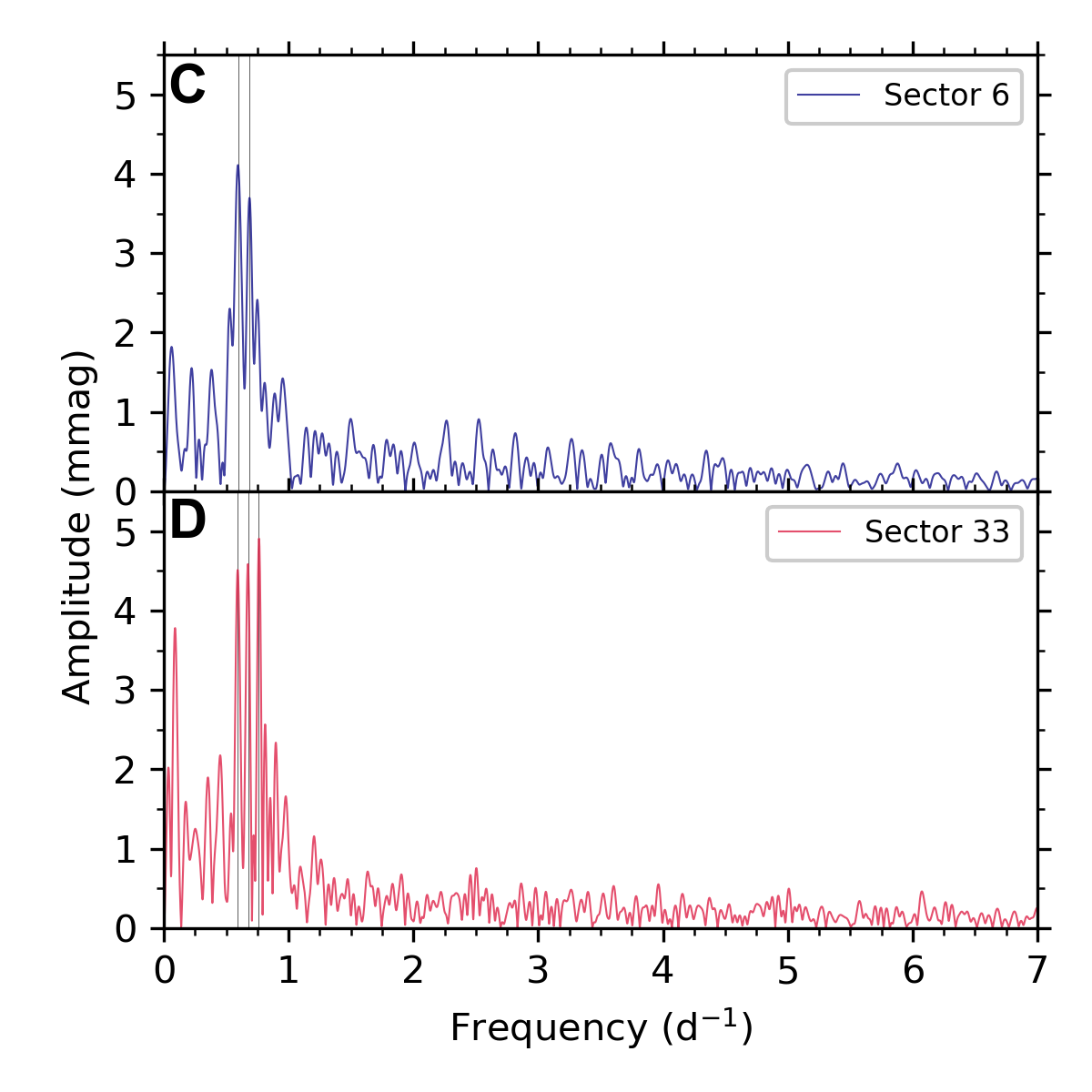}
    \caption{{\bf Frequency analysis of the TESS light curve of HD~45166}.  A and B: TESS light curve from sectors 6 (A) and 33 (B) of HD~45166, with relative brightness in units of millimagnitude (mmag) in the TESS passband $\Delta$Tp on the ordinate axis and time (in BJD) on the abscissa.  C and D: Amplitude spectra of sector 6 (C) and 33 (D) TESS light curves, with significant frequencies (i.e.\ exceeding the significance threshold S/N $>$ 4; see text for detailed) marked by vertical gray lines.  
    } 
    \label{fig:TESS}
\end{figure}

To analyze the light curve of HD~45166, we downloaded $15\times15$ pixel cutouts from the TESS full frame images (FFIs) stored at MAST using the {\scshape Astrocut} package \cite{ASTROCUT}  within the interactive {\sc lightkurve} software \cite{lightkurve2018}. We extracted light curves for HD~45166 by defining custom aperture masks, and estimated the background flux from the median flux per frame excluding pixels that contain stellar flux. We subtracted the background flux from the target flux, normalized by dividing through the median. A final linear trend was subtracted from each light curve to remove remaining instrumental effects. Our FFI-extracted light curves for sectors 6 and 33 are shown in the left panel of Fig.~\ref{fig:TESS}A and B.

To determine the variability in each light curve, we used Fourier analysis and the method of iterative pre-whitening \cite{Bowman2021c}. We fit the light curve of each sector with a model function of the form:

\begin{equation}
    \Delta m \left( t \right) = \sum_{i} A_i \cos \left( 2\,\pi\,\nu_i \left(t- t_{\rm ref}\right) + \phi_i\right),
\label{eq:Fourier}
\end{equation}
where $t$  is the time stamps of the light curve with units of BJD--2457000.0,  with BJD being the Barycentric Julian date (preferred over MJD for high-precision  space-mission photometry), $t_{\rm ref}$ is a reference time at the end of the light curve of each sector ($t_{\rm ref} = 1490.0$ and 2225.0 for sectors 16 and 33 in units of BJD-2457000.0, respectively),  and $\nu_i$, $A_i$, and $\phi_i$ are the  frequency, amplitude, and phase of the of the $i^{\rm th}$  cosinusoid term, respectively.

To determine the total number of significant frequencies, we use a significance criterion of an amplitude signal-to-noise ratio (S/N) larger than four in the amplitude spectrum, where the signal is defined as the amplitude of a peak and the noise is defined as the average amplitude in the residual amplitude spectrum in a symmetric window of width 1~d$^{-1}$ centered at the extracted peak for each iteration \cite{Bowman2021c}. In total, we detected two and three  frequencies fulfilling our significance criterion in the sectors 6 and 33 TESS data of HD~45166, respectively. We performed  an independent multi-frequency non-linear least-squares  fit to each light curve using Eq.\,(\ref{eq:Fourier}) to optimize the frequencies and their corresponding amplitudes and phases, and determine correlated uncertainties, which are provided in Table~\ref{tab:TESS}.

The shorter 10-min cadence of the sector 33 data yields a lower noise level in the amplitude spectrum, on average, because of the factor three larger number of data points compared to the sector 6 data. This contributes to the detection of the additional frequency in the sector 33 TESS data, and the overall smaller uncertainties for the optimized parameters in Table~\ref{tab:TESS}. Given the low frequency resolution of a short (27\,d) TESS light curve of $1/\Delta(t) \simeq 0.04$~d$^{-1}$, the two lower frequencies in sector 6 are the same as those in sector 33.

Multi-periodic variability in a period range of several days is typical of gravity mode pulsations in main-sequence B-type stars \cite{Bowman2020c}. The dominant frequency of $\approx 0.6\,{\rm d}^{-1}$ lies close to the period of $1.66\,$d detected from RV variations for the B-type component [\cite{Steiner2005}, see also below]. We infer that the 1.6\,d period, previously attributed to the orbital period of the system, is instead due to pulsations in the B7~V component. This is consistent with the much longer orbital period we find below.
% which lies in the 
% Periods in this range detected through spectral line profile variability have previously been attributed to the orbital motion of the stars in the HD~45166. However, here we conclusively demonstrate that such a period in fact originates from pulsations which supports the much longer orbital period of HD~45166 found in our spectroscopic analysis.

\begin{table}
\centering
% \normal
\caption{{\bf Significant frequencies extracted from the TESS light curves of HD~45166.} Frequencies exceeding the significance criterion  \cite{MaterialsMethods} in sectors 6 (upper part) and 33 (lower part) of the TESS light curve are listed in the first column. The amplitudes and phases are given in the second and third columns (Eq.\,\ref{eq:Fourier}). }
{\begin{tabular}{lccc}\hline \hline
\vspace{-2mm}\\ 
&   Frequency   &   Amplitude     &   Phase   \\
&   (d$^{-1}$)   &   (mmag)     &   (radian)   \\
\hline
Sector 6: \\
&   $0.6006 \pm 0.0012$     &   $4.23 \pm 0.20$    &   $1.618 \pm 0.094$    \\
&   $0.6825 \pm 0.0014$     &   $3.75 \pm 0.20$    &   $-0.11 \pm 0.11$   \\
   \hline
Sector 33: \\
&   $0.59401 \pm 0.00056$    &   $4.45 \pm 0.11$    &   $1.114 \pm 0.045$    \\
&   $0.67593 \pm 0.00057$    &   $4.46 \pm 0.12$    &   $1.622 \pm 0.045$    \\
&   $0.75981 \pm 0.00054$    &   $4.68 \pm 0.12$    &   $0.142 \pm 0.043$    \\
   \hline
\end{tabular}}
\label{tab:TESS}\end{table}

\subsection*{Revised orbital analysis of HD~45166}

To measure the RVs from the available spectra, we cross-correlate the observations with appropriate templates, following an established procedure \cite{Zucker1994}. Measuring the RVs of the B-type component is only possible with the FEROS, HERMES, and ESPaDOnS spectra given the lower wavelength coverage and quality of the LNA spectra. To form the template of the B7~V component, we first use the synthetic {\scshape powr} model (Fig.\,\ref{fig:Blines}). We then shift-and-add all observations using the measured RVs, and use this to produce a calibrated template of co-added observations \cite{Shenar2020LB1}. The new template is then cross-correlated with the observations to yield the final absolute RVs of the B7~V component. We use the strong absorption lines Si\,{\scshape ii}\,$\lambda \lambda 4128, 4131$ and  Mg\,{\scshape ii}\,$\lambda \lambda 4481.1, 4481.3$ for the cross-correlation, and find that our conclusions do not depend on the choice of lines.

\begin{figure}[!htb]
   \centering
\includegraphics[width=\textwidth]{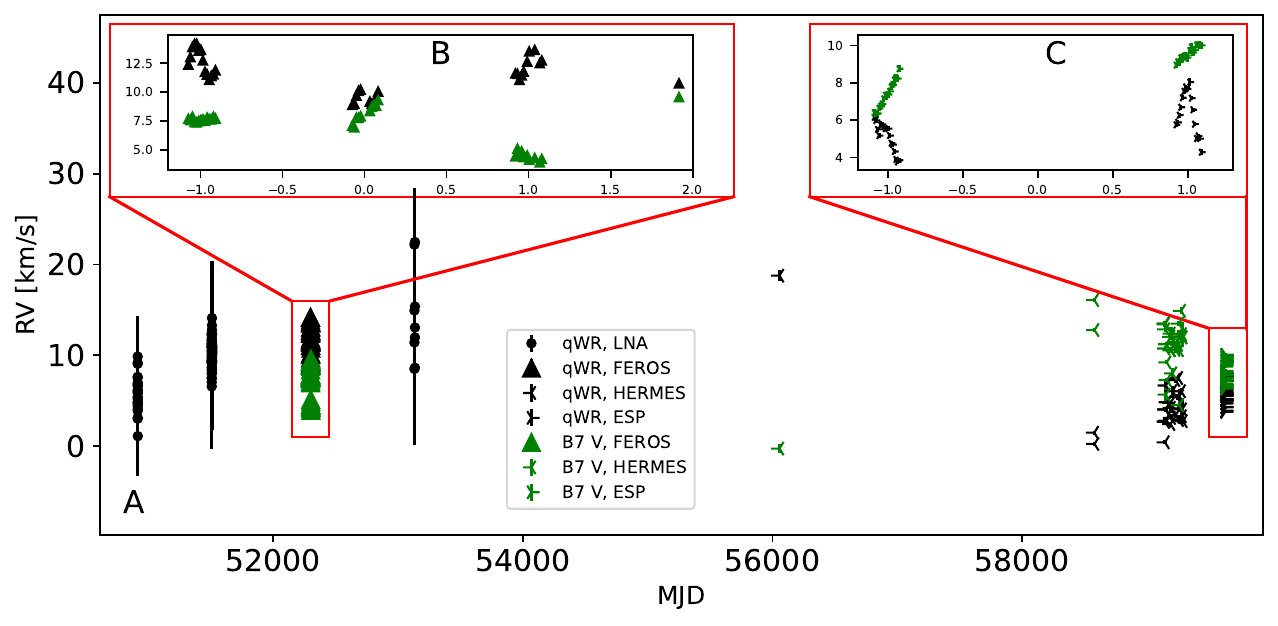}
    \caption{{\bf Overview of RVs measured for the two components of HD~45166}. (A) Black and green symbols show the 199 and 96 RVs derived for the Wolf-Rayet and B7~V components, respectively. Each instrument is marked by a different symbol (see legend). The relative RVs of the Wolf-Rayet component were vertically shifted to have the same mean as the absolute RVs of the B7~V component. (B and C) Short-term variability of both components. The origin of the $\approx 5\,$hr period of the Wolf-Rayet is uncertain, but could be related to pulsations. We argue (see above) that the $\approx 1.6\,$d period is due to pulsations in the B7~V component. There is a long-term anti-phase motion of the two components, indicating a period of $\approx 20$ years. Error bars depict formal 1$\sigma$ uncertainties (the uncertainties of HERMES and ESPaDOnS RVs are comparable to symbol sizes).
    } 
    \label{fig:ScatPlot}
\end{figure}

For the Wolf-Rayet component, we use a similar procedure. However, lacking a synthetic template, we use one of the FEROS observations as a first template. We use the emission-line complex in the range $4630 - 4660$\,\AA\ for cross-correlation. After measuring the preliminary RVs, which are measured relative to the observation used as a template, we form a co-added, high S/N template to remeasure the RVs. We allow for an overall systematic shift between the RVs of the Wolf-Rayet  and B7~V components, such that they result in the same systemic velocity (see below). The choice of emission lines used for cross-correlation does not affect our results, though usage of highly variable lines such as He\,{\scshape ii}\,$\lambda 4686$ yields a substantially larger scatter, and is therefore avoided. 
A total of 199 and 96 RVs for the Wolf-Rayet and B7~V components are available.   An overview of the derived RVs is shown in Fig.\,\ref{fig:ScatPlot}, and they are also provied as supplementary data (table S1).

\begin{figure}[!htb]
   \centering
\includegraphics[width=.65\textwidth]{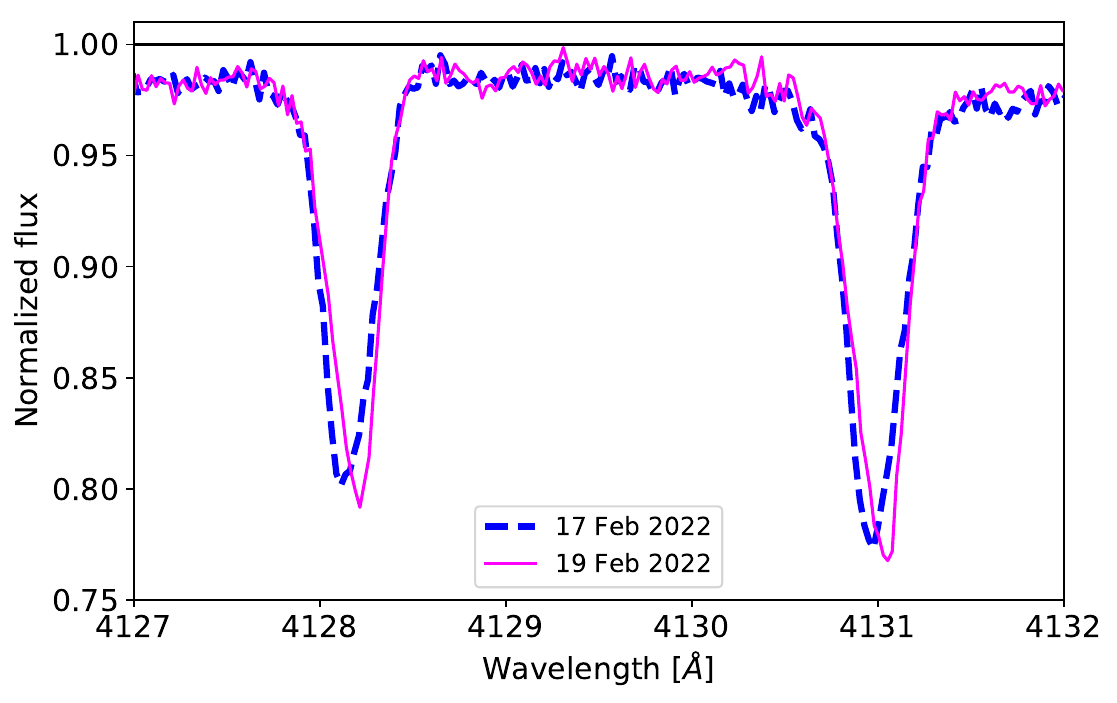}
    \caption{{\bf Spectroscopic signature of gravity-mode pulsations in the B7~V component.} Shown are co-added nightly ESPaDOnS spectra taken during 2022 Feb 17 and 19 Feb 2022, focusing on the Si\,{\scshape ii}\,$\lambda 4128, 4131$ doublet of the B7~V component. There are line profile variations, with their cores shifting in apparent RV, but their wings being static. This behavior is typical for non-radial gravity-mode pulsations \cite{Aerts2009}.
    } 
    \label{fig:Pulsations}
\end{figure}

Both components exhibit substantial variability on several timescales, and this variability appears to be associated with several periods, consistent with previous investigations \cite{Steiner2005}.
 The RVs of the B7~V component vary on a short timescale, and we can confirm the 1.6\,d period found previously \cite{Steiner2005} and formerly attributed   to the orbital period of the system. By subtracting several periods from the RVs measured for the Wolf-Rayet component, the 1.6\,d period was previously found \cite{Steiner2005} in the Wolf-Rayet component as well. However, having repeated this experiment using our dataset, we do not confirm this result. Instead, we find that the 1.6\,d period is associated with non-radial gravity-mode pulsations in the B7~V component (see above). This is evident in the ESPaDOnS spectra, which have pronounced variability between the two observing nights.  To boost the S/N, we co-added all observations acquired during a single night, and show the resulting spectra in Fig.\,\ref{fig:Pulsations}. This spectral variability is typical for non-radial gravity-mode pulsations in slowly rotating stars \cite{Aerts2009}.

 The strong emission lines of the Wolf-Rayet star display variability with a short period of the order of 5\,hr, as previously identified \cite{Steiner2005}. The origin of this period, which persist pseudo-periodically for at least 20\,yr, is not clear. However, this period cannot  correspond to orbital motion, since the associated RV semi-amplitude is $\approx$ 1 to 2\,$\kms$ and varies from epoch to epoch. The combination of a 5\,hr period with the low RV amplitude would either imply a highly unlikely pole-on geometry, or a sub-stellar mass for the companion, which is inconsistent with stellar evolution.  It is possible that the 5\,hr period is due to pulsations in the Wolf-Rayet component, in a phenomenon related to its magnetic field, or in an interplay between the two.
%  \textcolor{blue}{The 5\,hr period is also not readily evident in the TESS light curve. Hence, the available data do not allow us to evaluate this periodicity unambiguously. It may be originating from the Wolf-Rayet component, as it also seems to show longer-term RV changes that are of the order of days, but more detailed theoretical models are required to verify this.} %A possible origin for the 5\,hr period could be pulsations in the Wolf-Rayet component. Since the dominant mode of pulsations scales as the inverse of the root of the mean density $\rho^{-1/2}$, we can obtain a rough estimate for the pulsational period of the Wolf-Rayet component by comparing it to the 1.6\,d period of the B component, using the masses and radii provided in Table\,\ref{tab:Parameters}.  The values given in Table\,\ref{tab:Parameters} are  indicative of a pulsational period of $\approx 10 \pm 5\,$hr for the Wolf-Rayet component, consistent within  $1\sigma$ with the 5\,hr period. The Wolf-Rayet component also seems to show longer-term RV changes that are of the order of days. However, the data do not enable to constrain this periodicity unambiguously. We propose that these periods may be pulsations in the Wolf-Rayet component, though why the 5\,hr period is not readily evident in the TESS light curve is not clear.
 
 Upon inspection of the long-term variability of the RVs of both components, anti-phase motion between the two components is apparent. While the short-term variability partly masks this trend, we conclude that the two stars are bound on a long-period orbit. In the HERMES and ESPaDOnS observations taken during 2019-2022, the RVs of the B7~V component in Fig.\,\ref{fig:ScatPlot} are persistently red-shifted with respect to those of the Wolf-Rayet component. However, exactly the opposite is seen in the FEROS spectra, acquired in 2002. The individual HERMES observation taken in 2012 again shows a reversal of the RVs. 
 
 To obtain an orbital solution, we binned the RVs on a 30\,d baseline to smooth-out short-term variability in the two components. To account for the short-term variability, we attribute an additional uncertainty of $3\,\kms$ to the individual measurements (representative of the amplitude of the short-term RV variation), but divide this by the number of measurements per 30\,d bin, though we require the uncertainties to remain at least 1\,$\kms$. We fitted the binned RVs with a model of the RV curve resulting from the orbital parameters ($P, t_0, e, \omega, K_1, K_2, v_0$) using the Python lmfit package \cite{LMFIT, Shenar2019}. We also included a systematic shift between the RVs of the Wolf-Rayet component and a systematic shift of the LNA RVs (see above). The LNA offset  is found to be $-2.8\pm 1.6\,\kms$, which is small and does not impact our solution.  
 The solution is shown in Fig.\,\ref{fig:Orbit}, which corresponds to the parameters given in Table\,\ref{tab:Parameters}.   The derived period is 22.5\,yr, though this value depends  on the treatment of the uncertainties. The formal uncertainties on the orbital parameters are derived from the covariance matrix; a Markov Chain Monte Carlo (MCMC) simulation yielded similar uncertainties.

\begin{figure}[!htb]
   \centering
\includegraphics[width=.75\textwidth]{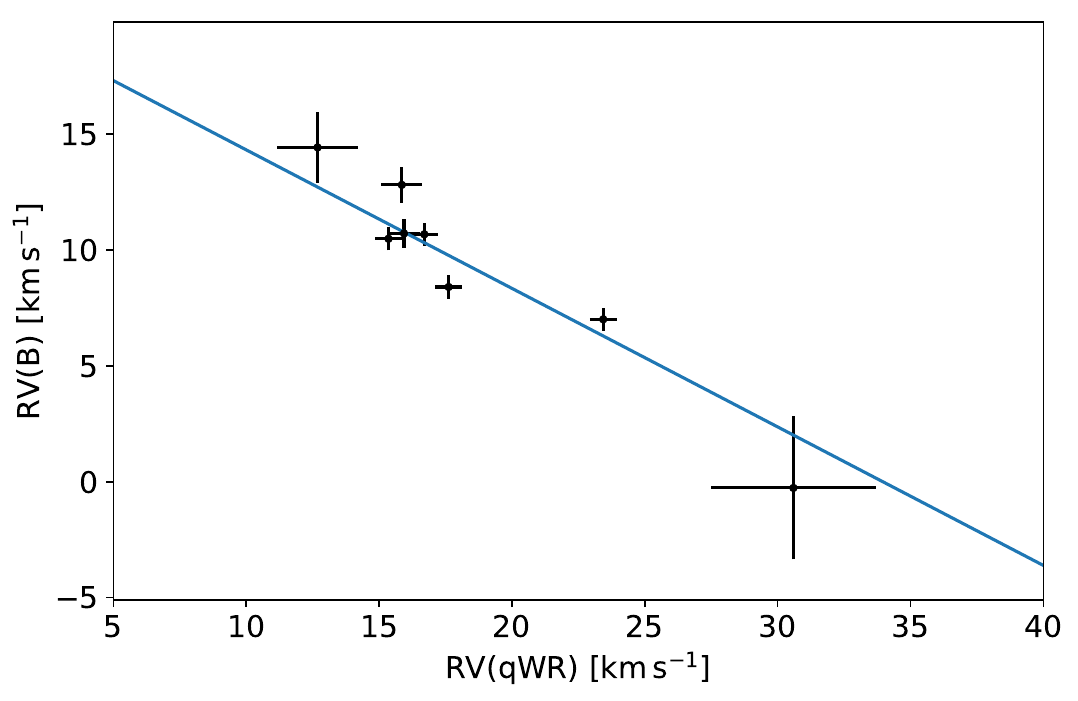}
    \caption{{\bf Derivation of the mass ratio.} The RVs of the B7~V component (y-axis) are shown as a function of those of the Wolf-Rayet component (x-axis). Measurements are the same as those used for the orbital analysis. The line is a linear regression to the data; its negative slope (mass ratio) and intercept are $q = 0.60\pm0.13$ and $C = 20.3\pm2.3$. } 
    \label{fig:linReg}
\end{figure}

While the ratio $K_2/K_1$ yields $q=0.58\pm0.16$,  we can determine the mass ratio independently of the orbital elements  by considering that, at each point of time, the RVs of the components scale in proportion to their semi-amplitudes \cite{Rauw2000}:

 \begin{equation}
     {\rm RV(B)} = -q\cdot {\rm RV(qWR)} + C,
\label{eq:linv}
 \end{equation}
where  RV(B) and RV(qWR) are the RVs of the B7~V and Wolf-Rayet components at a given epoch, respectively, $q \equiv M_{\rm qWR}/M_{\rm B}$ is the mass ratio, and C is a constant that depends on the systemic velocities and $q$. By performing a linear regression to the measurements (Fig.\,\ref{fig:linReg}), we find  $q = 0.60\pm0.13$, which agrees well with the value derived from the orbital analysis.

The evolutionary mass of the B7~V component, in combination with the mass ratio, leads to the mass of the Wolf-Rayet component.  Using our derived evolutionary mass of $M_B = 3.38\pm0.10$ and the derived  mass ratio of $q = 0.60\pm0.13$ yields $M_{\rm qWR} = 2.03\pm0.44\,M_\odot$. From $M_{\rm B} \sin^3 i = 1.50\pm0.74\,M_\odot$ and $M_B = 3.38\pm0.10$, we infer $i = 49\pm11^\circ$. The long period of the system provides an explanation for the low RV amplitudes observed in the system, without the need to invoke a pole-on configuration. We find that the Wolf-Rayet component exceeds the Chandrasekhar mass limit, and is expected to undergo core-collapse (see below).

\subsection*{Evolutionary scenario} 

Assuming that the Wolf-Rayet component formed through a merger of two helium stars (see main text),  we construct an evolutionary model for the system  using the \texttt{MESA} stellar evolution code \cite{Paxton2011}. We considered the evolution of a $5M_\odot$ star with a $3M_\odot$ companion at an initial orbital period of two days, representing the inner binary in an initial triple. The more massive component overflows its Roche lobe first, undergoing mass transfer both during and after its main sequence, resulting in a stripped $0.63M_\odot$ star. Mass transfer increases the orbital period and separation to $66$ days and $138R_\odot$, while the accretor mass increases to $7.4M_\odot$. Such post-interaction binaries are known to exist \cite{Gies1998, Shenar2020LB1, Bodensteiner2020}. Owing to its increase in mass, the accretor evolves faster, finishing its main sequence and initiating a phase of inverse mass transfer before the stripped star depletes core helium. Due to the extreme mass ratio, the mass-transfer is unstable, and the system evolves into a common-envelope phase at an age of $130$ Myrs. 
% This age is consistent within $1\sigma$ with the derived age for the B7~V component.}

The outcome of common-envelope phases is poorly understood in  binary evolution \cite{Ivanova+2013}. One-dimensional models can be used to assess whether a merger would occur, or whether the hydrogen envelope would be ejected before the helium cores would coalesce. We follow an energy prescription for common envelope evolution \cite{Webbink1984}. The binding energy of the envelope down to a given mass coordinate $m$, $E_\mathrm{bind}(m)$, can be computed as
\begin{eqnarray}
E_\mathrm{bind}(m) = \int_m^M\left(-\frac{Gm}{r(m)}+u(m)\right)dm,
\end{eqnarray}
where $M$ is the total mass of the star, $r(m)$ is the radius of a shell enclosing the mass $m$, $G$ is the gravitational constant, and $u$ is the specific internal energy of the star, for which we also consider the energy from hydrogen and helium recombination. The source of energy to unbind this material is the change in orbital energy. To estimate the orbital energy released through a merger we consider that all layers above $m$ are ejected from the donor, and compute the orbital separation $a_\mathrm{RLOF}$ at which the stripped star (with a radius of $0.18R_\odot$ at the onset of common-envelope) would overfill its Roche lobe.  The energy released through the merger, $E_\mathrm{merger}(m)$, is computed then from the orbital energy at this separation,
\begin{eqnarray}
E_\mathrm{merger}(m) = \frac{Gm M_\mathrm{stripped}}{2a_\mathrm{RLOF}(m)},\quad a_\mathrm{RLOF}(m) = \frac{0.18\; R_\odot}{f(M_\mathrm{stripped}/m)}
\end{eqnarray}
where $M_\mathrm{stripped}$ is the mass of the original primary after it has been stripped and
\begin{equation}
f(q)=\frac{0.49q^{2/3}}{0.6 q^{2/3}+\ln\left(1+q^{1/3}\right)}
\end{equation}
is an approximation of the Roche radius \cite{Eggleton1983}. The layers that are ejected through the common envelope phase are then determined from the condition $E_\mathrm{bind}(m)=\alpha_\mathrm{CE}E_\mathrm{merger}(m)$, where $\alpha_\mathrm{CE} \in [0,1]$ is the efficiency of converting the orbital energy into ejection of the material.

\begin{figure}[!htb]
   \centering
\includegraphics[width=\textwidth]{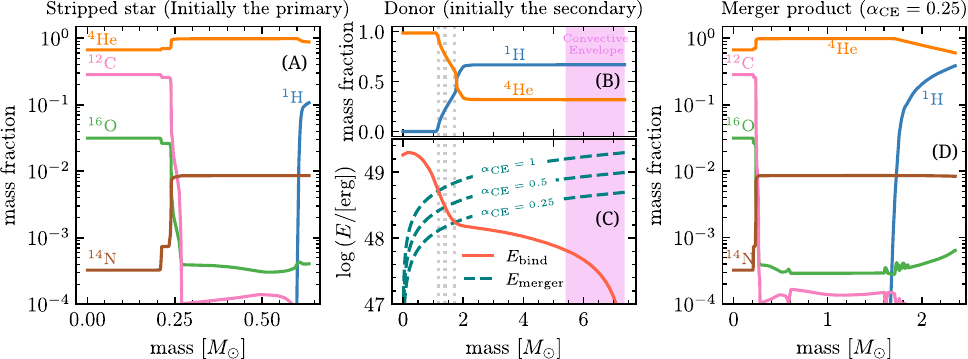}
    \caption{{\bf Abundance profiles and envelope binding energy of the pre- and post-merger product.} (A) abundance profile of the stripped star immediately before the merger.  (B, C) Abundance profile (B) and binding energy (C) of the donor star at the onset of the in-spiral, prior to envelope ejection.  Dashed teal curves in panel C show the energy that would be released through a merger after ejecting all material above a given mass coordinate for different efficiencies $\alpha_\mathrm{CE}$. The intersection between these curves with $E_{\rm bind}(m)$, denoted with dashed vertical lines, determines the point to which the donor star would be stripped. (D) abundance profile of the merger product under the assumption $\alpha_\mathrm{CE}=0.25$. Composition profile is constructed from both stars starting from the center with regions depleted in hydrogen in order of increasing helium abundance, and then layers with increasing hydrogen abundance.} 
    \label{fig:bind_en}
\end{figure}

The binding energy of the donor star and the energy that would be released through a merger after ejecting all material above a given mass coordinate are shown in Fig.\,\ref{fig:bind_en}. If orbital energy converts with full efficiency into removing the envelope ($\alpha_\mathrm{CE}=1$), only $0.08M_\odot$ of hydrogen rich material would remain, indicating that the system would eject the common envelope,  halting the in-spiral and merging process (Fig.\,\ref{fig:bind_en}). However, the actual efficiency is expected to be lower \cite{IaconiDeMarco2019}. Taking $\alpha_\mathrm{CE}=0.5$ or $0.25$ the system would retain $0.3M_\odot$ and $0.62M_\odot$ of hydrogen rich material, respectively. We perform a simple model of the post-merger evolution by constructing a star with a composition profile that combines that of the stripped star with the layers that are not ejected from the donor star (Fig.\,\ref{fig:bind_en}). The post-merger evolution depicted in Fig.\,\ref{fig:hr_merger} shows the evolution after the merged object reaches a state of thermal equilibrium as a core-helium burning object, until it reaches core helium depletion.

For the merger model with $\alpha_\mathrm{CE}=0.25$, we find the best match to the properties of the Wolf-Rayet component of HD 45166 occurs at has an age (starting from the initial binary model) of 133 Myrs, consistent (within $1\sigma$) with the age constraint provided by the B7~V component. The surface composition is rich in hydrogen, but due to the CNO cycle it has a very low carbon abundance ($X_C=6.5\times 10^{-5}$), which is inconsistent with previous measurements of the system \cite{Groh2008}. That analysis used a wind model that does not describe the magnetosphere (see above). However, the observed strength of carbon emission lines (Figs.\,\ref{fig:Spec} and \ref{fig:Zeeman}) in the spectrum indicates that the atmosphere of the Wolf-Rayet component contains a significant amount of carbon, which is inconsistent with our merger model. We argue that fallback from some of the outermost layers of the donor star could have contaminated the merger product with material that has not been CNO processed, leading to higher oxygen and carbon abundances observed at the surface in the present day.  While modeling the merger product beyond core-helium depletion is subject to various uncertainties [such as mass-loss in the presence of the magnetic field, \cite{Keszthelyi2019}], in our model, we find it forms a $1.29M_\odot$ O-Ne-Mg core with a $1.04M_\odot$ envelope composed primarily of helium and with a small amount ($0.04M_\odot$) of hydrogen. We do not model the evolution of the system beyond this point, but we expect the core would continue to grow from shell burning. We predict that the Wolf-Rayet component will become a supergiant ($R\approx 300\,R_\odot$, well within its Roche lobe) and explode as a type Ib or IIb electron-capture supernova  [\cite{Tauris2015}, their table 1], depending on the amount of hydrogen retained \cite{Dessart2011, Gilkis2022a}. The neutron star remnant is expected to possess a magnetic field strong enough to appear as a magnetar (see above).

\section*{Supplementary Text}

% \subsection*{Comparison to objects with similar spectra}

% Unique as it may be, HD~45166 is relatively nearby ($\approx 1\,$kpc away from Earth) and is sufficiently bright ($V= 9.88\,$mag) for other similar objects to have been potentially detected. From a statistical standpoint, {\bf there should be dozens of such objects {\bf within $\approx 5$\,kpc, depending on the volume distribution of these objects}. This is well within reach of various Galactic spectroscopic catalogs, such as that of Galactic Wolf-Rayet stars, which extends to $\approx 10\,$kpc and down to $V\approx 20\,$mag \cite{Rate2020}.} One may therefore wonder whether other ojects known in the literature have a similar nature to that of the Wolf-Rayet component in HD~45166. 

Parallels have been drawn between the spectrum of  HD~45166 and Wolf-Rayet stars belonging to the class of V~Sagittae stars \cite{Steiner1998}. This class comprises a few stars that portray Wolf-Rayet spectra with reported photometric and spectroscopic variability periods of the order of hours that are attributed to orbital periods. Specifically, WR~7a and WR~46 were proposed to exhibit properties similar to  HD~45166 \cite{Oliveira2003, Oliveira2004}.
However, both WR~7a and WR~46 exhibit broad, strong emission lines that resemble those of classical Wolf-Rayet stars, Moreover, no firm detection of a magnetic field could be obtained for WR~46 \cite{Hubrig2020}. Hence, while an investigation of the magnetic properties of WR~7a might still be worth pursuing, neither WR~7a nor WR~46  appear to be suitable candidates for massive magnetic helium stars. 

An intriguing similarity is observed when comparing the spectrum of the Wolf-Rayet component to a subset of central stars of planetary nebulae, classified as [WN] or [WC] stars \cite{Todt2010}. The spectral appearance and luminosities of some [WC] and [WN] stars matches well with those of the Wolf-Rayet component in HD~45166. However, a distinct difference is that HD~45166 has no nebula associated with it \cite{Willis1983}:  no evidence for nebulosity is seen in the spectrum, nor is it seen in  infrared images. While central stars of planetary nebulae are thought to represent the final evolutionary stages of solar-type stars following a ``standard'' post-asymptotic giant branch evolution, it is possible that a few impostors are hidden in this population. In the context of the evolutionary scenario proposed here, their nebulae could be the remnant of a violent merger event, implying that they are ``freshly born'' magnetic helium stars. Specifically, a few such objects were identified that share a similar spectral morphology to HD~45166, and notably the Of?p emission complex in the range 4630-4660\,\AA. Examples include  PB~8 \cite{Todt2010} and Hen~2-108. However, a campaign to measure the magnetic fields of a sample of central stars of planetary nebulae, including Hen~2-108,  did not find evidence for magnetism in these objects \cite{Steffen2014}. It remains to be seen whether other [WN] or [WC] stars exhibit strong magnetic fields.

\pagebreak
\newpage

\end{document}